%% file: main.tex
\documentclass[conference]{IEEEtran}

\usepackage{graphics}
\usepackage{fancyhdr}
\usepackage{xcolor}
\usepackage{tcolorbox}
\usepackage[hyphens]{url}
\usepackage{hyperref}
\usepackage{cite}
\usepackage{amsmath,amssymb,amsfonts}
\usepackage{algorithmic}
\usepackage{textcomp}
\usepackage{soul}
\usepackage{color}
\usepackage[framemethod=TikZ]{mdframed}
\usepackage[normalem]{ulem}
\usepackage{adjustbox}
\usepackage{xspace}
\usepackage{scalerel}
\usepackage{amssymb}
\usepackage{ctable}
\usepackage{pifont}
\usepackage{url}
\usepackage{mathtools}
\usepackage{tabularray}
\usepackage{multirow}
\usepackage{array}
\usepackage[normalem]{ulem}
\usepackage{caption}

\newcommand{\bhl}[1]{#1}
\newcommand{\cmark}{\ding{51}}%
\newcommand{\xmark}{\ding{55}}%
\newcommand{\defeq}{\vcentcolon=}
\newcommand\ourmodel{Carbon Responder\xspace}
\newcommand\meta{Meta\xspace}
\newcommand\ig{RTS1\xspace}
\newcommand\web{RTS2\xspace}
\newcommand\np{NP\xspace}
\newcommand\ppu{power usage\xspace}
\newcommand\pce{power capacity\xspace}

\DeclareCaptionFont{9pt}{\fontsize{9pt}{10pt}\selectfont}
\let\svthefootnote\thefootnote
\newcommand\blankfootnote[1]{%
  \let\thefootnote\relax\footnotetext{#1}%
  \let\thefootnote\svthefootnote%
}
\let\svfootnote\footnote
\renewcommand\footnote[2][?]{%
  \if\relax#1\relax%
    \blankfootnote{#2}%
  \else%
    \if?#1\svfootnote{#2}\else\svfootnote[#1]{#2}\fi%
  \fi
}
\title{Carbon Responder: Coordinating Demand Response for the Datacenter Fleet}

\author{
\IEEEauthorblockN{Jiali Xing$^*$, Bilge Acun$^\dagger$, Aditya Sundarrajan$^\dagger$, David Brooks$^\ddagger$,\\ Manoj Chakkaravarthy$^\dagger$, Nikky Avila$^\dagger$, Carole-Jean Wu$^\dagger$, Benjamin C. Lee$^*$ \\}\\
\IEEEauthorblockA{$^\dagger$Meta, $^*$University of Pennsylvania, $^\ddagger$Harvard University
}
}
\begin{document}
\maketitle
\pagestyle{plain}

\newcommand{\hpcaheight}{0mm}
\ifdefined\eaopen
\renewcommand{\hpcaheight}{12mm}
\fi

\footnote[]{Correspondence to: \textit{$<$xjiali@seas.upenn.edu$>$,~$<$acun@meta.com$>$}}

\input{00_abstract}

\input{01_intro}

\input{02_dc_dr}
\input{02_cr_per_se}
\input{03_cost_models}
\input{04_response_policy}

\input{05_results}
\input{07_relatedwork}

\input{08_discussion}

\section*{Acknowledgements}
We would like to thank Leonardo Piga and Hong-Shuo Chen from Meta for production workload characterization and their feedback on the project.
We also thank Kim Hazelwood for supporting this work.


\bibliographystyle{IEEEtranS}
\bibliography{CarbonFreeDatacenter}

\end{document}

%% file: 00_abstract.tex
\begin{abstract}

The increasing integration of renewable energy sources results in fluctuations in carbon intensity throughout the day. To mitigate their carbon footprint, datacenters can implement demand response (DR) by adjusting their load based on grid signals. However, this presents challenges for private datacenters with diverse workloads and services. One of the key challenges is efficiently and fairly allocating power curtailment across different workloads. In response to these challenges, we propose the \ourmodel framework.

The \ourmodel framework aims to reduce the carbon footprint of heterogeneous workloads in datacenters by modulating their power usage. Unlike previous studies, \ourmodel considers both online and batch workloads with different service level objectives and develops accurate performance models to achieve performance-aware power allocation. The framework supports three alternative policies: Efficient DR, Fair and Centralized DR, and Fair and Decentralized DR. We evaluate \ourmodel polices using production workload traces from a private hyperscale datacenter. Our experimental results demonstrate that the efficient \ourmodel policy reduces the carbon footprint by around 2x as much compared to baseline approaches adapted from existing methods. The fair \ourmodel policies distribute the performance penalties and carbon reduction responsibility fairly among workloads.

\end{abstract}

%% file: 01_intro.tex
\section{Introduction}
\label{sec:intro}
Hyperscale datacenters consumed tens of terawatt hours of energy in 2022~\cite{google21, meta21, microsoft21, amazon21}. Energy consumption for technology companies, such as Google and Meta, doubled from 2017 to 2020~\cite{google21, meta21}. \bhl{This rapid growth has motivated datacenters to reduce their operational carbon with \textit{supply-side solutions} that emphasize clean energy supply.} They have invested in renewable energy generation and storage to offset datacenter consumption as well as developed renewable energy contracts and credits to track those offsets~\cite{google13:ppa, facebook21:tariffs}. \bhl{However, supply-side solutions incur \textit{embodied carbon costs}: the carbon footprint from manufacturing  wind/solar farms and batteries.} These solutions become prohibitively expensive when datacenters must compute through periods of scarce renewable energy supply from intermittent sources such as wind and solar. For example, datacenters may need to increase wind and solar investments by an additional $5\times$ to increase the percentage of hourly carbon-free compute from 95\% to 99\% than from 0\% to 95\%~\cite{acun23}.

More effective solutions must coordinate supply and demand, adjusting datacenter activity in response to the energy grid's carbon intensity, i.e. implement \textit{demand response (DR)}. DR in the context of datacenters means deferring computation or degrading quality-of-service when carbon intensity is high. Equally important, it boosts computation when carbon intensity is low, ensuring deferred tasks dequeue rather than accumulate across time. 
%
Such load shifting can effectively reduce carbon emissions by leveraging the significant variation in a power grid's carbon intensity. 

\begin{figure}
	\centering
	\includegraphics[width=\columnwidth]{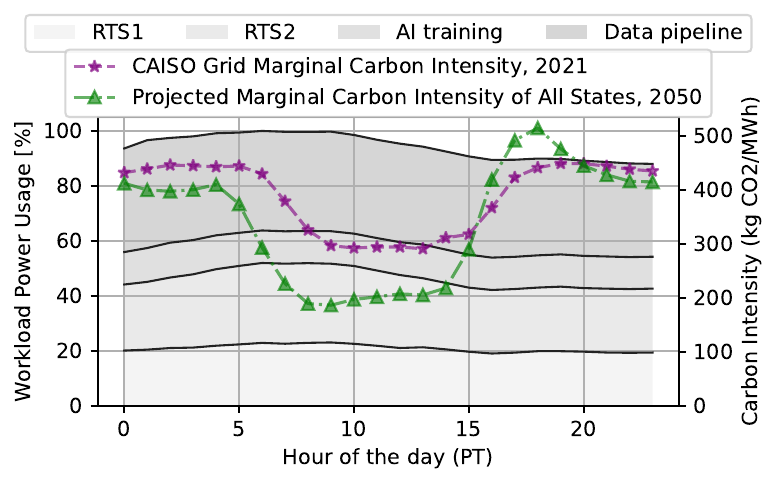}
  	\vspace*{-.25in}
   \caption{The plot demonstrates a notable increase in the variation of carbon intensity of the grid over time, indicating greater potential for demand response. The plot also shows the power usage breakdown of four example services: real-time services (RTS), AI, and data pipeline.}
    \vspace{-.2in}
	\label{fig:dr}
\end{figure}

For example, Figure~\ref{fig:dr} presents the normalized power usage of a cluster consisting of four workloads and illustrates the marginal carbon intensity\footnote{We choose the projected carbon intensity of all States for illustration, while the predicted carbon intensity of California shows the same variation.} based on the California grid (CAISO) data in 2021 \cite{watttime_mer} and the projected trend for 2050 \cite{ gagnon_cambium_2023}. 
\textit{Marginal carbon intensity} is the carbon footprint of the power plant at the margin of the grid's dispatch stack\bhl{: If electricity demand increases, the marginal power plant increase generation and, if demand falls, it would be the first plant to reduce generation.} 
Figure 1 indicates the peak-to-trough difference in marginal carbon intensity is significant. The trough can be as low as 66\% of the peak in today’s grid. Moreover, because fluctuations in carbon intensity are anticipated to increase, the trough can be as low as 40\% of the peak by 2050 \cite{gagnon2022long}.
Another analysis projects even greater growth in solar energy supply, leading to periods of zero marginal carbon intensity by 2050~\cite{eia_outlook_23}. 
\bhl{Furthermore, in today's grids substantial renewable energy generated is curtailed -- i.e. goes to waste. For instance, in California in 2022, 29 million megawatts were curtailed, amounting to 4.4\% of solar and wind generation} \cite{california_iso}. \bhl{In China 15\% of renewable energy was curtailed in 2019} \cite{bunodiere_renewable_2020}. \bhl{With increased renewable adoption, oversupplies issues due to the intermittent nature of renewable energy are expected to be more frequent.}
This trend underscores the immense potential for datacenters to reduce carbon and utilize the renewable energy better through demand response.

Despite its benefits, demand response is particularly challenging when the datacenter fleet supports a diverse mix of batch and realtime workloads --- \textit{Which workloads should respond when carbon-free energy is scarce, by how much, and when?} To address these questions, we propose a framework --- \ourmodel (CR) --- that integrates representative workload models with expressive demand response policies. CR analyzes the impact of demand response for individual workloads and unifies these impacts into a common measure of performance loss. This permits CR to compare performance costs and sustainability benefits from power adjustments across workloads. 
We develop DR policies for both realtime and batch workloads with varying \bhl{Service Level Objectives (SLOs\footnote{ \bhl{An SLO specifies the deadline by which a batch job should be completed.}})}, accounting for their relative sensitivity to power allocation based on production datacenter traces. In contrast, prior DR studies focus exclusively on batch workloads and make simplifying assumptions about performance loss (\textit{e.g.}, 20\% of power is deferrable within a 24-hour period without any loss) \cite{acun23, radovanovic_carbon-aware_2021, krioukov_design_2012}.

Moreover, we propose a family of sustainability-aware policies for CR. Subject to operational carbon targets, policies differ in their objectives. Some seek efficiency by minimizing the datacenter's aggregate performance loss when adjusting power (\textit{Efficient DR}) whereas others seek fairness by balancing losses across workloads based on their performance models (\textit{Fair DR}). Policies also differ in their implementation. Centralized policies formulate an optimization to be solved whereas distributed policies formulate mechanisms to incentivize participation from selfish agents. We explore the policy space and assess sustainability, performance efficiency, and fairness trade-offs.

Using \ourmodel, we evaluate a variety of demand response policies with production workload traces from a private hyperscale datacenter. Our experimental results demonstrate that the three CR policies exhibit an inherent trade-off between efficiency and fairness. When compared to baseline approaches adapted from existing works, the efficient \ourmodel policy achieves a carbon footprint reduction of 1.5x to 2x given the same performance degradation. In addition, the fair \ourmodel policies distribute the performance penalties and carbon reduction responsibility more fairly among workloads than most baselines.

%% file: 02_dc_dr.tex
\section{Datacenter Demand Response}
\label{sec:background}

\input{baseline_table}

\begin{figure}
		\centering
	\includegraphics[width=1\columnwidth]{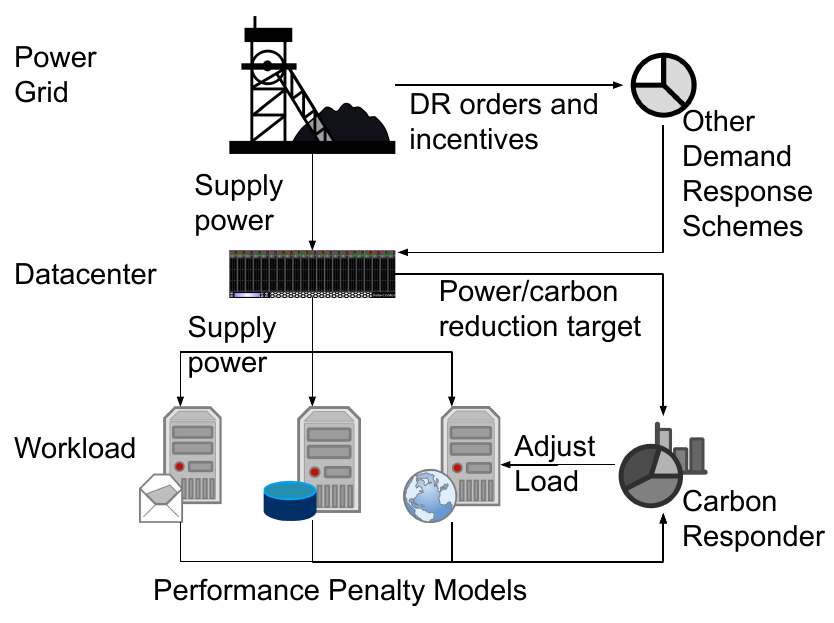}
    \vspace{-15pt}
	\caption{{\ourmodel} responds to grid's carbon signals by altering individual workload's power demand while prior work only modulates aggregate demand.}
 \vspace{-.2in}
	\label{fig:cr}
\end{figure}

\textbf{DR Abstraction Layers}. Demand response requires well defined interfaces between the grid, datacenter, and workloads as illustrated in Figure~\ref{fig:cr}. At the grid-datacenter interface, the grid supplies power for the datacenter’s hundreds of thousands of machines. The grid also provides signals about its supply and carbon intensity (\textit{e.g.}, energy prices, curtailment requests), which could permit intelligent datacenter demand response. At the datacenter-workload interface, the datacenter sets a power or carbon reduction objective and achieves it by scheduling workloads and modulating their energy demands.

Many prior demand response studies focus on the grid-datacenter interface \cite{liu_pricing_2014, acun_carbon_2022, radovanovic_carbon-aware_2021}. 
They treat the datacenter as a large consumer and assume some power usage can be deferred without penalty. Other DR papers, in contrast, schedule batch jobs directly in response to grid signals, neglecting or breaking abstraction layers \cite{goiri_greenslot_2011, goiri_greenhadoop_2012, liu_renewable_2012, zhang_hpc_2022}. The abstractions and assumptions of those work hinder their use for datacenters with heterogeneous users and workloads. 

Datacenter DR must consider several dimensions of the management problem. First, DR must define an optimization objective that formalizes the outcome sought when curtailing power. The outcome may be some combination of reductions in energy cost, peak power, and operational carbon. Second, DR must implement an allocation procedure that determines how power curtailments are distributed across heterogenous workloads. Finally, DR may wish to account for fairness and the contributions of individual workloads toward the datacenter’s broader sustainability goal.

\subsection{Objective -- Datacenter Demand Response for Carbon}

Table~\ref{tab:baselines} compares related work with our proposed \ourmodel framework. Studies such as \cite{liu_pricing_2014, zhang_hpc_2022} focus on demand response programs provided by power grids for grid reliability. On the other hand, \cite{meisner_power_2011, govindan_benefits_2011, wu_dynamo_2016} explore datacenter power capping, which has potential implications for datacenter carbon reduction. Note that datacenter DR goes beyond traditional power capping by rescheduling computation to periods when carbon intensity is low. 

There exists significant opportunity for carbon-informed datacenter demand response. Figure~\ref{fig:dr} indicates that grid carbon intensity varies significantly while datacenter power usage stays relatively stable. Modern datacenters exhibit little hourly variation in energy usage because they schedule computation to maximize utilization of their installed compute capacity and infrastructure \cite{fan_power_2007}. These time series suggest DR must re-discover time-varying demands for servers and align that demand with the grid’s renewable energy supply.

\subsection{Allocation -- Apportioning Power Adjustments}

In hyperscale datacenters, apportioning power adjustments to heterogeneous and diverse workloads is challenging. Prior studies avoid this challenge by focusing DR on a single workload or category of workloads. Many focus on batch workloads ~\cite{radovanovic_carbon-aware_2021, zhang_hpc_2022, acun_holistic_2022, liu_renewable_2012} while others focus on real-time workloads~\cite{govindan_benefits_2011, wu_dynamo_2016}. However, this narrow focus is neither sufficient nor efficient for several reasons. 

First, no single class of workloads can adjust enough power to align datacenter demand with fluctuations in energy supply and carbon intensity. When most of the datacenter's power is attributed to online workloads, modulating only delay-tolerant, batch workloads would be insufficient for sustainability and incur prohibitive perforamnce losses.
For example, 30-40\% of Google's workloads are delay-tolerant with a 24-hour SLO~\cite{tirmazi_borg_2020} and 20-30\% of Meta's are delay-tolerant with varying SLOs~\cite{acun_carbon_2022}. Although 70\% of Microsoft's Azure workloads are labeled delay-tolerant, the degree of tolerance is unspecified~\cite{cortez_resource_2017}.

\subsection{Fairness and Incentives}

Fairness and incentives play a crucial role in the allocation of power adjustments across workloads, particularly in private datacenters where teams have their own capacity entitlements and dedicated job scheduling frameworks. 
The studies discussed thus far neglect fairness, in part, because they lack performance models for heterogeneous workloads, cannot assess performance outcomes, and cannot quantify associated fairness implications \cite{radovanovic_carbon-aware_2021}. Thus, they could not assess the performance outcomes and the associated fairness implications.

Incentives may be required to encourage teams and their workloads to adjust power.
However, research on DR in private hyperscale datacenters has largely overlooked the role of incentives, despite some ideas in using markets to allocate system resources \cite{chase01, guevara13, zahedi18}.
Prior studies neglect incentives, in part, because they assume a centralized scheduler can compel DR within the datacenter, as exemplified by Google's use of Borg to the number of available CPUs when carbon intensity is high \cite{radovanovic_carbon-aware_2021}. 

%% file: baseline_table.tex

\begin{table*}[ht]
\centering
\resizebox{\textwidth}{!}{%
\begin{tabular}{l|ll|lll|l}
\hline
\multirow{2}{*}{Related work} &
  \multicolumn{2}{c|}{Objective} &
  \multicolumn{3}{c|}{Allocation} &
  \multirow{2}{*}{\begin{tabular}[c]{@{}l@{}}Providing \\ Fairness \\ Options\end{tabular}} \\ \cline{2-6}
 &
  \multicolumn{1}{c|}{Problem statement} &
  \begin{tabular}[c]{@{}c@{}}Optimization \\ Metric\end{tabular} &
  \multicolumn{1}{c|}{\begin{tabular}[c]{@{}c@{}}Workload \\ Type \end{tabular}} &
  \multicolumn{1}{c|}{\begin{tabular}[c]{@{}c@{}}Across-workload \\ Apportion Strategy\end{tabular}} &
  \begin{tabular}[c]{@{}c@{}}Model \\ Performance \\ Impact\end{tabular} &
   \\ \specialrule{.12em}{.05em}{.05em}  OLDI
\cite{meisner_power_2011} &
  \multicolumn{1}{l|}{DC Power efficiency} &
  Proportionality &
  \multicolumn{1}{l|}{Realtime} &
  \multicolumn{1}{l|}{Even split} &
  \cmark &
  \xmark \\ \hline eBuff
\cite{govindan_benefits_2011} &
  \multicolumn{1}{l|}{DC Power capping} &
  Electricity cost &
  \multicolumn{1}{l|}{Realtime} &
  \multicolumn{1}{l|}{Not applicable} &
  \cmark &
  \xmark \\ \hline Dynamo
 \cite{wu_dynamo_2016} &
  \multicolumn{1}{l|}{DC Power capping} &
  Peak power &
  \multicolumn{1}{l|}{Realtime} &
  \multicolumn{1}{l|}{Priority rank based on perf impace} &
  \xmark &
  \xmark \\ \hline Pricing DR
 \cite{liu_pricing_2014} &
  \multicolumn{1}{l|}{Grid-level DR} &
  Competitive ratio &
  \multicolumn{1}{l|}{Batch} &
  \multicolumn{1}{l|}{Not applicable} &
  \xmark &
  \cmark \\ \hline AQA
  \cite{zhang_hpc_2022} &
  \multicolumn{1}{l|}{Job-level DR} &
  Electricity cost &
  \multicolumn{1}{l|}{Batch} &
  \multicolumn{1}{l|}{Not applicable} &
  \cmark &
  \xmark \\ \hline Google 
 \cite{radovanovic_carbon-aware_2021} &
  \multicolumn{1}{l|}{DC DR} &
  Carbon and peak &
  \multicolumn{1}{l|}{Batch} &
  \multicolumn{1}{l|}{Priority tiers based on SLOs} &
  \xmark &
  \xmark \\ \hline
 Our model &
  \multicolumn{1}{l|}{DC DR} &
  Carbon and perf &
  \multicolumn{1}{l|}{Batch \& Realtime} &
  \multicolumn{1}{l|}{Optimization based apportioning} &
  \cmark &
  \cmark \\ \hline
\end{tabular}%
}
\caption{Related work}
\vspace{-.2in}
\label{tab:baselines}
\end{table*}

%% file: 02_cr_per_se.tex
\section{\ourmodel}
\label{sec:design} 

To achieve efficient and fair carbon reduction in private hyperscale datacenters, we propose 
\ourmodel(CR) --- a carbon-based demand response framework that focuses on the datacenter-workload interface in Figure~\ref{fig:cr}.
\bhl{Based on the current practices of our hyperscale datacenter, CR specifically focus on the operational carbon emissions attributed to the datacenter's consumption of grid power, assuming no batteries, or on-site renewable energy generation} \cite{govindan_benefits_2011, deng_concentrating_2011}.
It treats the datacenter as a collection of heterogeneous workloads and designs DR based on the performance characteristics of those workloads, rather than modeling the datacenter as a monolithic consumer \cite{liu_pricing_2014, acun_carbon_2022, radovanovic_carbon-aware_2021}. 
Furthermore, CR shields users and their workloads from the grid’s complexity and departs from prior studies that neglect or break these abstractions \cite{buchanan_carbon-aware_2023, goiri_greenslot_2011, goiri_greenhadoop_2012, liu_renewable_2012}.


\subsection{Framework Design}

\begin{figure}
	\centering
	\includegraphics[width=1\columnwidth]{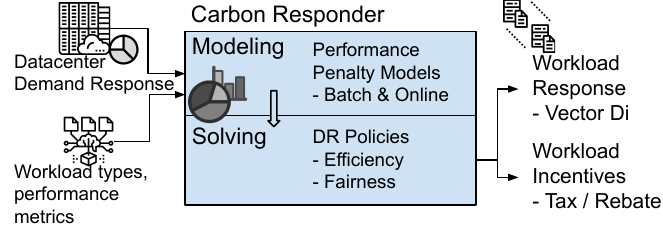}
         \vspace{-10pt}
	\caption{{\ourmodel} takes as input the datacenter DR and workload information to determine service response and incentives.}
        \vspace{-.2in}
	\label{fig:crs}
\end{figure}

CR recognizes that modulating datacenter power requires models of performance-power trade-offs and policies governing efficiency-fairness trade-offs. Figure~\ref{fig:crs} illustrates its two major contributions. First, CR trains models that quantify penalties when implementing demand response for diverse workloads. These models capture the relationship between power allocation and diverse measures of performance and service quality. Power allocations affect processor utilization, which in turn affect performance. Performance metrics vary by workload type (\textit{i.e.}, batch or real-time). \ourmodel aggregates individual workload characteristics to model the relationship between power and performance for the entire datacenter. 

Second, CR balances penalties incurred against carbon reduced when determining how much each workload should contribute towards the datacenter’s DR objective. \ourmodel optimizes how workloads modulate power use, minimizing performance penalties across all workloads. \ourmodel provides different policies to trade-off efficiency and fairness. 


\ourmodel takes as input workloads' measurements, which detail power usage and performance outcomes, and the datacenter's DR objectives. It learns the \textit{Performance Penalty Function} (Section~\ref{sec:penalty}) given the behavior of each workload's scheduler. Based on the penalty function and a fairness policy, \ourmodel outputs optimized power adjustments for each workload and details methods to enforce them (Section~\ref{sec:policies}). The adjustment determines each workload's performance penalty, power usage, and contribution to operational carbon. \bhl{Load adjustments are executed daily, while the CR performance modeling pipeline can be scheduled weekly or only subsequent to significant application updates to accommodate any application changes.}

\subsection{Supporting Diverse Workloads}

\begin{figure}
	\centering
    \includegraphics[width=1\columnwidth]{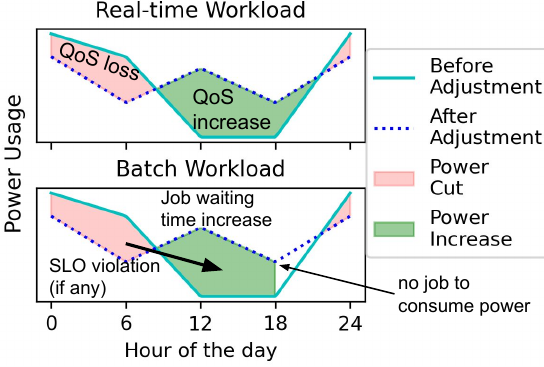}
   	\vspace{-.2in}
	\caption{Illustration of workload types and their characteristics.}
	\label{fig:penalty}
\end{figure}

\ourmodel extracts flexibility from both batch and real-time workloads, unlike previously proposed solutions. For each additional unit of power required for DR adjustment, \ourmodel identifies the workload with the smallest marginal performance penalty. When \ourmodel adjusts power usage, different types of workloads are affected differently as illustrated in Figure~\ref{fig:penalty}. 

\textbf{Real-time.} For real-time workloads, increasing power curtailments may degrade quality of the service and harm user engagement. In contrast, decreasing curtailments can improve service as workloads opportunistically exploit additional compute. 

\textbf{Batch without SLOs.} Batch jobs are often assumed to be delay-tolerant with no penalty. For example, AI training must be completed but do not have a strict deadline. For batch jobs without SLOs, \ourmodel models the DR penalty as the job's total waiting time \cite{graham_optimization_1979}. 

\ourmodel ensures the queue of deferred jobs does not accumulate across multiple days. If $d_t$ is the workload adjustment for hour $t$, then $\sum_{t{\in}Day} d_{t} = 0$. Figure~\ref{fig:penalty} shows how jobs deferred during DR (red, $d_t<0$) are rescheduled later (green, $d_t>0$) such that jobs complete within the day. 

\textbf{Batch with SLOs.} In production systems, many batch jobs specify landing times. These jobs are often part of a data analysis pipeline and their completion times affect downstream jobs. Penalties arise when completion times extend beyond the landing time and violate the SLO. We measure penalties in terms of \textit{tardiness}, the amount of extra time required beyond the landing time. Figure~\ref{fig:penalty} shows that DR causes some deferred jobs to incur a tardiness penalty while others might meet their SLOs. {\ourmodel} models total performance penalty associated with DR as a function of the red area.


\ourmodel's approach to workload elasticity is novel. Prior studies often take a macro view of workloads and make simplifying assumptions. Studies with batch jobs often assume they can be deferred arbitrarily and without penalty as long as jobs complete within 24 hours\cite{liu_pricing_2014, radovanovic_carbon-aware_2021}, neglecting penalties from violating SLOs and landing times. Other studies with real-time jobs assume that SLOs can be relaxed by some percentage \cite{wu_dynamo_2016, govindan_benefits_2011, krioukov_design_2012}, again neglecting penalties. Studies that model performance and power trade-offs as we do \cite{he_zeta_2012, reddy_optimizing_2017, chen_quantifying_2006}, envision neither DR nor coordination between real-time and batch services.  

%% file: 03_cost_models.tex
\section{Workload Performance and Penalty Models}
\label{sec:penalty}

\begin{table}[]
\centering
\resizebox{\columnwidth}{!}{%
\begin{tabular}{l|l|l|l}
\hline
\bf Services &
  \bf Description &
  \bf Category &
  \bf SLO \\ \specialrule{.12em}{.05em}{.05em} 
\begin{tabular}[c]{@{}l@{}}Data \\ Pipeline\end{tabular} &
  \begin{tabular}[c]{@{}l@{}}Storing, processing and \\ querying data. \end{tabular} &
  Batch &
  \begin{tabular}[c]{@{}l@{}}Five tiers of \\ SLOs: 1,2,4,8 \\ and $\infty$ hours.\end{tabular} \\ \hline
\begin{tabular}[c]{@{}l@{}}AI \\ Training\end{tabular} &
  \begin{tabular}[c]{@{}l@{}}AI model training \\ for production.\end{tabular} &
  Batch &
  No SLO \\ \hline
\ig &
  \begin{tabular}[c]{@{}l@{}}Serving real-time \\ requests for \ig app.\end{tabular} &
  Realtime &
  QoS based \\ \hline
\web &
  \begin{tabular}[c]{@{}l@{}}Serving real-time \\ requests for \web app.\end{tabular} &
  Realtime &
  QoS based \\ \hline
\end{tabular}%
}
\caption{Four representative workloads used in modeling and experiments.}
\vspace{-.2in}
\label{tab:workload}
\end{table}
 

\begin{table*}[]
\centering
\resizebox{\textwidth}{!}{%

\begin{tabular}{l|l}
\hline
\bf Required Input & \bf Source of Input \\ \specialrule{.12em}{.05em}{.05em} 
Power usage of each service before load adjustment & Daily average of \ppu from production workloads in hyperscale datacenters. \\ \hline
Performance penalty functions of online workloads & 
Adapted from the results of performance profiling in Dynamo  (\S~\ref{sec:realtime})~\cite{wu_dynamo_2016}. \\ \hline
Performance penalty functions of batch workloads  & Machine learning on production workload traces
(\S~\ref{sec:batch}). \\ \hline
Weights in penalty functions & Aligning performance penalty with entitlement loss when capping 15\% capacity (\S~\ref{sec:output}).    \\ \hline
Marginal carbon intensity of grid power & Daily average from CAISO 2021 (kg CO2/MWh) obtained from WattTime~\cite{watttime_mer}. \\ \hline
\end{tabular}%
}
\caption{Data Provenance}
\vspace{-.1in}
\label{tab:eva}
\end{table*}

\textbf{Model Input.} \label{sec:input} The model's input is a vector of hourly adjustments to power load based on one metric and two concepts: \textit{Normalized Power (\np)}, \textit{Power Capacity Entitlement}, and \textit{Physical Power Usage}. 
\np is the unit of power used by datacenter provisioning teams in place of Watts.
Power capacity is the workload's maximum permissible power usage and represents an entitlement to computational resources. 
Power usage is the workload's actual power usage.

Let $\vec U_i=[U_{i,1}, U_{i,2}, \cdots, U_{i,t},\cdots)$ denote \ppu for workload $i$ across time, and $\vec d_i=[d_{i,1}, d_{i,2}, \cdots, d_{i,t},\cdots)$ denote load curtailment vector measured in terms of \ppu. Positive $d_{i,t}$ indicates a load decrease (\textit{e.g.}, $d_{i,t} = 5$ means workload $i$ reduces load by 5 \np at time $t$) whereas negative $d_{i,t}$ indicates a load increase. Adjustments $d$ are relative to baseline load such that $d_{i,t}$ is the difference in \ppu with and without DR.





\textbf{Model Output.} \label{sec:output} The model's output is the penalty that arises from hourly adjustments to power load. \ourmodel first uses machine learning to calculate DR's performance loss and then scales performance loss into an equivalent loss in \pce. 

The first step accounts for unique, workload-specific measures of performance. The second step establishes a datacenter-wide measure of penalty, permitting comparisons across heterogeneous workloads. It models a linear relationship between losses in \pce and performance, calculating the scaling weight $k_i$ as the \pce loss divided by the performance loss.

\subsection{Modeling Heterogeneous Workloads}
\label{sec:data}


Without loss of generality, we illustrate and apply \ourmodel to model four \meta services --- Data Pipeline, AI Training, and two different real-time services (\ig, \web)---as detailed in Table~\ref{tab:workload}. These four services include the largest and most representative workloads in batch and user-facing computation. 

Table~\ref{tab:eva} details our data sources and inputs. Power usage data is obtained from a trace of production workloads in a hyperscale datacenter. The trace details daily average power usage over the year 2021, while the job-level traces for AI training and storage comprise 10,000 jobs subsampled within a two-day window. To model batch workload performance, we utilized production data. To model batch workloads, we use production data. To model real-time workloads, we use published Dynamo parameters \cite{wu_dynamo_2016} rather than our own experiments to ensure confidentiality. 


\subsubsection{Real-Time Workloads}
\label{sec:realtime}
\ig and \web represent \textit{real-time} services that must generate timely responses for users. Although real-time services cannot defer their computation, their quality-of-service (QoS) can be reduced in exchange for less power consumption \cite{krioukov_design_2012}. 
Dynamo profiles the effect of power capping on web server performance and finds latency is an increasing convex function of the power reduction. Based on Dynamo’s Figure 13 \cite{wu_dynamo_2016}, we fit a polynomial function for latency degradation $f = a_3 \delta_{it}^3 + a_2 \delta_{it}^2 + a_1 \delta_{it} $ where $ \delta_{it} = \frac {d_{it} } {U_{it} } \times 100 $ is the power adjustment expressed as a percentage of usage.

We fit distinct models $f_{\ig}$ and $f_{\web}$ based on published median and maximum latency degradation, respectively. These latency models are used to assess penalty $C_i$ from power adjustments $\vec{d}_{i}$. Thus, penalty functions for real-time workloads are:
\begin{equation}
\label{eq:realcost}
	C_i(\vec d_i) 
    = \sum_t k_i \times f_i(\delta_{it}) ,\
    \delta_{it} = \frac {d_{it} } {U_{it} \times 100} 
\end{equation}
where $f_{\ig} = 6.3 \delta_{it}^3 - 13 \delta_{it}^2 + 51.6 \delta_{it} 
    $ and $
    f_{\web} = -4 \delta_{it}^3 - 3.5 \delta_{it}^2 + 42.5 \delta_{it}$, and the weights $\{k_\text{\ig}, k_\text{\web}\}$ are calculated with the methodology specified in the 4$^{th}$ paragraph in Section~\ref{sec:output}.

\subsubsection{Batch Workloads}
\label{sec:batch}
Data Pipeline represents \textit{batch workloads with SLOs}. These workloads consist of data processing jobs that are critical to other services. \bhl{Five priority tiers correspond to five SLOs with deadlines of $[1, 2, 4, 8, +\infty]$ hours.} 
AI Training represents \textit{batch workloads without SLOs}. These workloads consist of offline training jobs that run within a capacity allocation and without an explicit deadline. 

The penalty function for batch services captures how DR lengthens waiting time and induces tardiness. We model waiting time and tardiness as a function of DR adjustments, specifying a regression model with engineered features and fitting that model with Lasso regression\footnote{\bhl{Lasso (Least Absolute Shrinkage and Selection Operator) is a regression method that includes feature selection and regularization.}} and cross-validation. We obtain training data by implementing a scheduler, simulating schedules under varied processor availabilities, and measuring tardiness. We implement an earliest due date (EDD) scheduler, but \ourmodel supports any scheduling framework.


\textbf{Power Adjustment $\rightarrow$ Scheduling Results.} First, a linear model estimates the processor availabilities (CPUs/GPUs) based on the power supply.
Then, we implement an earliest due date (EDD) scheduler to simulate how processor availabilities influence batch job performance. 
The simulator's inputs include hourly energy capacity, server capacity, and a trace of batch jobs. 
The simulator reports waiting time and tardiness --- the waiting time beyond what can be tolerated by the SLO for each job \cite{koulamas_total_1994}.






\textbf{Scheduling Results $\rightarrow$ Machine Learning Model.} The model's dependent variable is tardiness and waiting time for batch jobs with and without SLOs, respectively. Modeling waiting time directly as a function of load adjustment $\vec d_i$ is a naive starting point, producing the simple penalty function:
$ c_i ( \vec d_i ) =\sum_{t'}^T \beta_{t'} \times d_{t'}+\beta_0.$
However, this model uses too many features. And it neglects the cumulative impact of curtailed power; jobs delayed in previous hours are queued which lengthen the waiting time in every hour they remain queued. This naive model fits poorly and we can do better with engineered features. 

\input{feature_table}



To incorporate more meaningful features into our machine learning model, we derive the total waiting time and its variations as independent variables. These features are outlined in Table~\ref{tab:features}, where $x^+$ denotes the positive part of $x$, given by $x^+ = \max \{x,0\}$, and $|{J}_{i,t}|$ represents the total number of jobs for workload $i$ at time $t$.

The first feature we investigate is the cumulative waiting time of all jobs, as indicated in the first row of Table~\ref{tab:features}. The terms in the inner sum estimate the number of jobs that are delayed due to power adjustments $d_{i,t'}$ at time $t'$. The inner sum calculates the number of queued jobs at time $t$ by accumulating the delayed jobs from all previous hours. Taking the positive part of the sum ensures that the queue length is non-negative. The outer sum aggregates the queue length per hour over all hours $T$ in the schedule, providing a measure of the total waiting time.

Similarly, the second feature examines cumulative waiting time with regard to power usage (\np) instead of the number of jobs. This feature quantifies the total delayed power usage in units of ($\np \cdot \text{hour}$), while the first feature is measured in units of ($\text{job} \cdot \text{hour}$).
The third feature considers the convex relationship between power cut and waiting time. To capture this convexity, we introduce a squared term as a potential feature.
The fourth feature estimates the total number of jobs affected by power curtailment, offering a non-cumulative measure to assess the impact of demand response.
The final feature accounts for tardiness and represents the number of jobs queued for more than a specified Service Level Objective (SLO) threshold (SLO hours). It quantifies the total overdue hours for jobs that have waited more than the SLO threshold.

We collect training data by generating diverse curtailment vectors $d$, scheduling jobs from \meta, and measuring the model's inputs (features) and outputs (tardiness). Diverse curtailments are sampled with a random walk \cite{klusacek_scheduling_2021}, using only those where average curtailment is positive. 

We train the machine learning model using Lasso. 
It regularizes when fitting coefficients,  using hyperparameter $\alpha$ to balance minimizing residuals and constraining the magnitude of the coefficients; $\alpha$ is set with ten-fold cross-validation. 

Table~\ref{tab:features} specifies features selected for batch workloads. We specify separate models with different variables for batch workloads with SLOs (Data pipeline) and without (AI training). The dependent variable is tardiness and waiting time for batch jobs with and without SLOs, respectively. 



\begin{table}[]
	\centering
	\resizebox{\columnwidth}{!}{%
	\begin{tabular}{l|l|l|lll}
\hline
\multirow{2}{*}{\textbf{Workload}} & \multirow{2}{*}{\textbf{\# Samples}} & \multirow{2}{*}{\begin{tabular}[c]{@{}l@{}}\textbf{\# Features}\\ \textbf{selected}\end{tabular}} & \multicolumn{3}{l}{\textbf{10-Fold Cross Validation}}                                                                                                                            \\ \cline{4-6} 
                          &                                                     &                                                                                & \multicolumn{1}{l|}{\begin{tabular}[c]{@{}l@{}}\textbf{MAE}\\ \textbf{Mean}\end{tabular}} & \multicolumn{1}{l|}{\begin{tabular}[c]{@{}l@{}}\textbf{MAE}\\ \textbf{Var.}\end{tabular}} & \textbf{R2}    \\ \specialrule{.12em}{.1em}{.1em}
AI Training               & 303                        & 2                                                                              & \multicolumn{1}{l|}{150.0}                                             & \multicolumn{1}{l|}{24.7}                                              & 0.789 \\ \hline
Data pipeline                   & 162                                             & 2                                                                              & \multicolumn{1}{l|}{39.2}                                              & \multicolumn{1}{l|}{14.3}                                              & 0.864 \\ \hline
\end{tabular}%
	}
\caption{LASSO regression accurately learns performance penalties as a function of load adjustment for batch workloads.}
\label{tab:lasso}
\end{table}

\begin{figure}[!htbp]
		\centering		
 	\vspace{-.2in}
		\includegraphics[width=0.9\columnwidth]{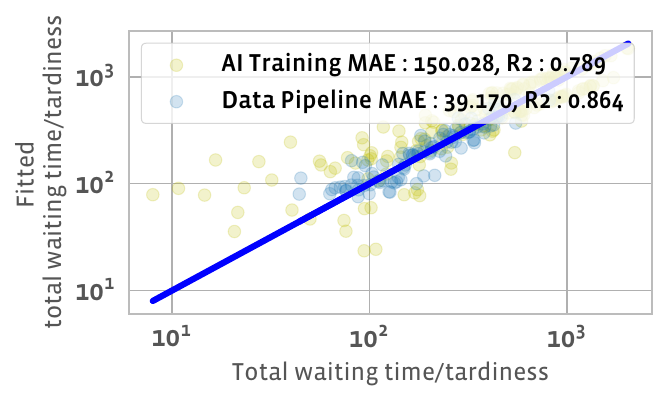} 
 	\vspace{-.1in}
		\caption{LASSO regression accurately predicts performance penalties as a function for batch workloads.}
		\label{fig:both_lasso}
\end{figure}

The regression effectively models batch workload's performance penalty as a function of adjustment. Table~\ref{tab:lasso} demonstrates a good model fit. Figure~\ref{fig:both_lasso} illustrates accurate predictions, plotting fitted penalties against measured ones.
With  $x_1, \cdots, x_2$ specified in Table~\ref{tab:features}, penalty functions for batch workloads are: 
\begin{align}
\label{eq:batchcost}
	C_i ( \vec d_i ) & = \left( k_i ( \beta_0 + \beta_1 x_1 + \beta_2 x_2 )\right)^{+} 
\end{align}

\begin{figure}[ht]
	\centering
	\includegraphics[trim={1cm 0cm 1.5cm 1cm},width=0.6\columnwidth]{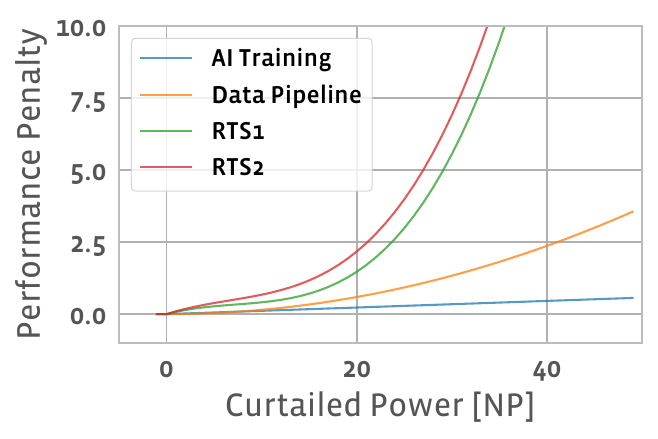}
 	\vspace*{-.1in}
	\caption{Penalty functions of different services}
        \vspace{-.2in}
	\label{fig:costexample}
\end{figure}

%% file: feature_table.tex
\begin{table}[]
\centering
\resizebox{\columnwidth}{!}{%
\begin{tabular}{l|l|l|l}
\hline
\textbf{Feature} & \textbf{Equation} & \begin{tabular}[c]{@{}l@{}}\textbf{Selected} \\ \textbf{for AI} \end{tabular} & \begin{tabular}[c]{@{}l@{}}\textbf{Selected} \\ \textbf{for Data} \\ \textbf{pipeline} \end{tabular} \\ \specialrule{.12em}{.1em}{.1em} 
\begin{tabular}[c]{@{}l@{}}Waiting time\\ (jobs) \end{tabular} & $\sum_{t=0}^{T} \left(\sum_{t'=0}^{t} |{J}_{i,t'}| \cdot \frac{d_{i,t'}}{U_{i,t'}}\right)^+$      & \xmark & \xmark \\ \hline
\begin{tabular}[c]{@{}l@{}}Waiting time\\ (power) \end{tabular} & $\sum_{t=0}^{T} \left(\sum_{t'=0}^{t} d_{i,t'} \right)^+$                                         & \cmark \quad \bhl{as $x_1$} & \cmark \quad \bhl{as $x_1$}  \\ \hline
\begin{tabular}[c]{@{}l@{}}Waiting time\\ squared \end{tabular} & $\sum_{t=0}^{T} \left(\sum_{t'=0}^{t} |{J}_{i,t'}| \cdot \frac{d_{i,t'}^2}{U_{i,t'}}\right)^+$    & \xmark & \cmark \quad \bhl{as $x_2$}  \\ \hline
\begin{tabular}[c]{@{}l@{}}Number of\\ jobs delayed \end{tabular}  & $ \sum_{t'=0}^{T} \left( |{J}_{i,t'}| \cdot \frac{d_{i,t'}^+}{U_{i,t'}}\right)$                   & \cmark \quad \bhl{as $x_2$}  & \xmark \\ \hline
\begin{tabular}[c]{@{}l@{}}Total\\ tardiness \end{tabular}       & $\sum_{t=0}^{T} \left(\sum_{t'=0}^{t-SLOs} |{J}_{i,t'}| \cdot \frac{d_{i,t'}}{U_{i,t'}}\right)^+$ & N/A   & \xmark \\ \hline
\end{tabular}%
}
\caption{Derived analytical features and feature selection.}
\vspace{-.2in}
\label{tab:features}
\end{table}

%% file: 04_response_policy.tex
\section{Datacenter Demand Response Policies}
\label{sec:policies}

\ourmodel models heterogeneous workloads and enables DR policy, which apportions power adjustments across heterogeneous workloads. We propose several policy options that differ in optimization objectives and efficiency-fairness trade-offs. Moreover, we compare these policies against prior studies that have taken varied perspectives on datacenter DR, which are detailed in Table~\ref{tab:baselines}. These prior studies are not directly comparable because they target unique characteristics of different abstraction layers, datacenters, and workloads. Nonetheless, where possible, we adapt these policies and formulate the corresponding DR optimization rigorously.

\subsection{Carbon Responder Policies}

\ourmodel supports three policies that balance efficiency and fairness differently. The precise nature of this balance determines how \ourmodel formulates and optimizes power adjustments. The formulation, in turn, determines how the datacenter enforces or incentivizes users to participate in DR and contribute to power adjustments. 

For demand response, the decision variable is a matrix of hourly power adjustments $\mathbf{D} = [ \vec d_1, \vec d_2, \cdots, \vec d_i, \cdots, \vec{d}_W]$ for each workload $i \in W$. Let $C(\mathbf{D})=\sum_i C_i(\vec{d}_i)$ denote aggregate performance losses and $CF(\mathbf{D})$ denote the change in operational carbon footprint, which corresponds to the inner product of marginal carbon intensity and power adjustment). 

\textbf{CR1 -- Efficient DR} apportions datacenter-wide power adjustments $\mathbf{D}$ across workloads to minimize performance costs $C$ and carbon footprint $CF$. Hyperparameter $\lambda$ supports varied performance and carbon trade-offs. Smaller values of $\lambda$ emphasize carbon reductions at the expense of performance, producing larger power adjustments.
\begin{equation}
	\label{eq:wo}
	\min_{\mathbf{D}} \ \lambda \ C(\mathbf{D}) + CF(\mathbf{D}) 
\end{equation}
Efficient DR may induce unfairness as workloads with greater power efficiency, and thus smaller performance losses from DR, will experience larger curtailments. To enhance fairness, we consider two additional policies. 

\textbf{CR2 -- Fair and Centralized DR} ensures each workload makes an equal contribution to datacenter-wide power adjustments. When workloads all cap power by the same percentage, denoted by cap\%, each workload will suffer individual performance losses $C_i(\text{cap}\%)$. The CR2 policy minimizes operational carbon while ensuring performance loss $C_i(\vec{d}_i)$ for each workload $i$ is consistent with the loss from equal power caps $C_i(\text{cap}\%)$. 
\begin{equation}
	\label{eq:bdr}
    \min_{\mathbf{D}} \  CF(\mathbf{D}) ~~~~~ \text{s.t.} \ C_i (\vec d_{i}) = C_i({\text{cap}\%}), \ \forall i 
\end{equation}
CR2 uses equal power caps as a reference for fairness but does not actually cap power. Instead, CR2 determines power adjustments that minimize carbon subject to performance constraints that are deemed fair under hypothetical, equal power caps. It adjusts power allocations for individual workloads independently until their performance losses equal those associated with power capping at $\text{cap}\%$. This approach is preferable to simply equalizing losses across workloads (\textit{i.e.}, $C_i = C_j$), which can vary significantly in scale.

Both CR1 and CR2 rely on centralized enforcement. Each workload must implement prescribed adjustments through workload-specific hardware or software mechanisms. Non-compliance leads to an indiscriminate reduction in \pce, which subsequently decreases \ppu and penalizes performance. This enforcement mechanism ensures workloads adhere to the DR plans.




\textbf{CR3 -- Fair and Decentralized DR} encourages participation in power adjustments with decentralized implementation, using taxes and rebates, rather than centralized enforcement. First, the policy ensures initial taxes are collected fairly (\textit{i.e.} with same tax rate) across workloads. Then, the policy offers rebates to workloads that reduce carbon through DR. For each workload, a tax reduces its \pce whereas rebates offset the tax. Users and their workloads are motivated to earn rebates, which increase their final power allocations. Each user has an equal opportunity to earn rebates by adjusting power usage. 



Using taxes and rebates to reduce and increase \pce, respectively, can be formulated as decentralized optimization. Let $E_{i}$ denote workload $i$'s initial power capacity entitlement. Let $T_{i}$ and $P_{i}$ denote tax paid and rebate received, respectively. The workload's net power entitlement after adjusting for taxes and rebates is thus ${E}_{i} - T_{i} + P_{i}$. \ourmodel imposes three constraints on CR3's power optimization. 

First, we constrain actual hourly usage $\vec U_i$ such that it does not exceed workload $i$'s net entitlement. Larger rebate $P_i$ relaxes power constraints imposed by tax $T_i$, thereby improving performance. 
\begin{equation}
\label{eq:pce}
    \max ( U_{i} - \vec d_{i} ) \leq E_i - T_{i} + P_{i}
\end{equation}
We also ensure fiscal balance such that rebates offered to workloads are covered by taxes collected across workloads. Thus, the policy does not create or require extra power capacity.
\begin{equation} 
\label{eq:fiscal}
    \sum_{i \in W} P_{i} \leq \sum_{i \in W} T_{i}
\end{equation}
Finally, for fairness, each workload is taxed equally at the beginning but receives differentiated rebates based on its contribution to DR. 
\begin{equation}
\label{eq:equaltax}
    T_{i} = T_{j}, \ \forall i, j \in W.
\end{equation}

For example, in Optimization~\ref{eq:bilevel}, workloads are taxed and must relinquish some percentage of their initial power capacity entitlements (\textit{e.g.}, 20\% of $E_i$). Workloads are then offered rebates based on their participation in DR and contribution to carbon reductions (\textit{i.e.}, $P_i(\vec d_i) = CF(\vec d_i)$). With decentralized optimization, each workload determines its $\vec d_i$ in $\mathbf{D}$ to minimize its performance degradation $C_i$ subject to \ref{eq:pce}--\ref{eq:equaltax}.
\begin{equation}
	\begin{aligned}
	\label{eq:bilevel}
		\min_{\vec d_i} \ C_i (\vec d_i) ~~~ & \text{s.t.} \ P_i(\vec d_i) = CF(\vec d_i), T_i = 0.2\ {E}_i, 
	\end{aligned}
\end{equation} 
As a workload contributes more to DR, it earns a larger rebate that relaxes power constraints and improves performance. On the other hand, as the workload contributes less to DR, rebates that permit power usage in other hours of the day may be insufficient to offset performance losses. Thus, a workload should optimally increase DR contributions until these effects balance and the marginal increase in performance from a marginal increase in DR is zero.

\subsection{Baseline Policies}

Our baselines are derived from notable prior research, which encompass different design options. B1 and B2 represent simple and optimized power capping mechanisms, respectively \cite{govindan_benefits_2011}. B3 and B4 distinguish between diverse workload types, using either heuristics or optimization to reduce carbon. 
These power capping baselines, adapted from Meta and Google's studies \cite{wu_dynamo_2016, radovanovic_carbon-aware_2021}, were originally intended for reducing power costs and mitigating power emergencies but can be used to reduce carbon as well. 



\textbf{B1 -- Proportional Power Capping} reduces datacenter power usage by setting power caps or limits $L_i$ as a fraction $F$ of their \pce $E_i$ and ensuring this fraction is equal for all workloads such that $ \frac{L_i} {E_i} = \frac{L_j} {E_j} = F, \ \forall \ i, j \in W $ 
This policy calculates hourly adjustments to power usage $\vec U_i$ so that each workload conforms to its cap.
\begin{equation}
\label{equ:capadj}
\vec d_{it} = \max \{ \vec U_{i} - L_i, ~ 0 \}
\end{equation}
We sweep hyperparameter $F$ to quantify this policy's performance and carbon trade-offs. As power capping becomes more aggressive, the policy will produce larger performance penalties and carbon reductions.

\textbf{B2 -- Performant Power Capping} reduces datacenter power to minimize combined peak usage and performance loss. Peak power usage after DR is $\max_t \sum_{i \in W} ( \vec U_i - \vec d_i ) $ because $ \vec U_i - \vec d_i $ is workload $i$'s hourly power usage. This policy uses workload models $C$ from Section~\ref{sec:penalty} to optimally set differentiated power caps for each workload as follows $$\min_{\mathbf{D}} \ \left[\lambda \ C(\mathbf{D}) + \max_t  \sum_{i \in W} ( \vec U_i - \vec d_i ) \right] $$ 
Note that $\mathbf{D}$ is a matrix of power adjustments for workload $i$ and hour $t$. We sweep hyperparameter $\lambda$ to quantify this policy's performance and carbon-trade-offs. As $\lambda$ increases, the policy favors performance at the expense of power and carbon reductions. 

This policy is inspired by eBuff \cite{govindan_benefits_2011}, which shaves power peaks with a policy that balances reductions in electricity bills with losses in performance. Note that eBuff's policy is applied to one workload at a time and lacks a strategy for apportioning power curtailments to a mix of heterogeneous workloads. 
   

\textbf{B3 -- Prioritized Power Capping} employs priority-based heuristics for power capping. It protects batch workloads and curtails only real-time workloads, reducing their power usage based on a pre-defined priority order while ensuring reductions never exceed a pre-defined maximum cut. The priority order and maximum cut for each real-time workload is established by a human operator. 

For instance, suppose real-time workload $i$ has higher priority than $j$ and the maximum allowable power cut for both is 20\%. B3 will first curtail $j$'s power usage down to 80\% of its power cap and then curtail $i$'s. Curtailments are determined as earlier in Equation~\ref{equ:capadj}. As $i$ and $j$ are curtailed, performance deteriorates and carbon decreases, creating a trade-off. 

This strategy is derived from Dynamo \cite{wu_dynamo_2016}, which sets varying power caps for each front-end cluster based on a priority order of services. Interestingly, Dynamo finds that real-time workloads experience negligible performance degradation when servers are subject to power capping. This finding is explained by over-provisioned power buffers for real-time workloads, which ensures constant, high-quality service \cite{wu_dynamo_2016}. This finding motivates a priority order that caps real-time workloads and trims buffers before capping batch workloads. 


\textbf{B4 -- Load Shaping} schedules workloads to balance performance and carbon. It protects real-time workloads and curtails only batch workloads while ensuring their SLOs. This policy minimizes the weighted sum of carbon and daily peak power as follows. $$\min_{\mathbf{D}} \left[ CF(\mathbf{D}) + \lambda  \max_t  \sum_{i \in W} ( \vec U_i - \vec d_i ) \right]  \ \text{s.t. batch SLOs}$$ We sweep hyperparameter $\lambda$ to quantify this policy's performance and carbon-trade-offs. As $\lambda$ increases, the policy favors performance at the expense of power and carbon reductions. 

Such a policy is similar to that in Google's study of datacenter DR~\cite{radovanovic_carbon-aware_2021}, which assumes real-time workloads have low tolerance for power capping \cite{radovanovic_carbon-aware_2021}, protects them from demand response, and predominantly adjusts power usage for batch workloads. Other power oversubscription and capping schemes from Google \cite{sakalkar_data_2020} and Microsoft \cite{kumbhare_prediction-based_2021} suggest capping only non-production, non-critical workloads, which would produce power caps similar to B4's.

\subsection{Constraints}
In solving the associated optimization problem, both \ourmodel and the baseline policies adhere to two constraints: the total capacity constraint and batch preservation.

Firstly, the result of Demand Response (DR) should not exceed the datacenter's total power capacity. If it does, the DR policy would necessitate additional machines, thereby leading to an increased embodied carbon footprint, which would undermine the DR's objectives. Based on observations from our hyperscale datacenter practices, we assume that the datacenter maintains a 20\% buffer capacity that remains unused. Consequently, the total capacity is set to 120\% of the peak power usage, derived from the summation of power usage across the four workloads. Mathematically, this ensures that the peak power post-DR remains within the total capacity:
\begin{equation}
    \max_t \sum_{i \in W} ( \vec U_i - \vec d_i ) \leq 1.2 \sum_i E_i 
\label{eq:pcec}
\end{equation}

Secondly, both \ourmodel and the baseline policies ensure batch preservation. The policies reschedule hourly power usage but adjustments, for each batch workload, must sum to a non-negative value over the hours in a day.This is crucial to ensure that any delays experienced by batch workloads are not cumulative across days. Regardless of whether SLOs are adhered to or violated, this constraint ensures that batch jobs are executed to completion without indefinite delays.
\begin{equation}
    \sum_{t} ( \vec d_i ) \geq 0, \ \forall i \in \text{batch workload}
\end{equation}

%% file: 05_results.tex
\section{Experimental Evaluation}
\label{sec:experiment}

We evaluate DR policies, quantifying their ability to reduce carbon and examining the fairness of performance losses. First, we show DR dynamics by illustrating {\ourmodel}'s load adjustments during a representative day and detailing how rescheduling power usage reduces carbon and impacts performance (\S\ref{sec:dynamics}). Then, we compare DR policies and show {\ourmodel} is more efficient than baselines, achieving greater carbon reductions with smaller performance penalties (\S\ref{sec:efficiency}). Finally, we compare DR policies and show {\ourmodel} is more fair than baselines based on the dispersion of carbon reductions and performance penalties across workloads (\S\ref{sec:perworkload}-\ref{sec:fairness}).

\subsection{Experimental Methods}


First, \ourmodel models performance penalties resulting from power adjustments for each workload using equations \ref{eq:realcost} and \ref{eq:batchcost}. We use traces and data from four production workloads detailed in \S\ref{sec:data}. 

Second, we implement DR policies with optimization to determine load adjustment $\mathbf{D}$. We solve optimization problems with Scipy's Sequential Least Squares Programming. Optimization determines power adjustments for a two-day interval, allowing us to determine whether a policy delays batch jobs beyond 24 hours. Optimization is offline and day-ahead, which aligns with methods in prior work \cite{radovanovic_carbon-aware_2021}. We limit curtailments to at most half the workload's original power capacity entitlement, which accounts for the fact that idle power typically constitutes half of a server's total power usage and {\ourmodel} does not presently power down servers. 

Finally, we calculate carbon reductions for the two-day interval as the inner product of vectors for marginal carbon intensity and power reductions (\S\ref{sec:policies}). We process CAISO data to quantify marginal and average carbon intensity \cite{watttime_mer}. Carbon Responder can support the analysis of other balancing authorities and geographical locations by drawing on other EIA data \cite{eia_outlook_23}. We assess \ourmodel against baseline policies based on their carbon reduction and performance losses. Carbon reduction is normalized by the total operational carbon, while the performance losses are measured by the percentage of equivalent power capacity losses. 

The datacenter’s baseline power capacity is at the scale of many tens of megawatts. And the datacenter’s baseline operational carbon is at the scale of thousands of metric tons of CO2 for the two-day interval. For confidentiality, our evaluation reports carbon and power reductions as a percentage of baselines without DR. Percentages range from 1\% to 8\%, which correspond to carbon and power reductions at the scale of tens of metric tons of $CO_2$ and multiple megawatts, respectively.


\subsection{Carbon and Power Dynamics}
\label{sec:dynamics}
\begin{figure}
	\centering 
 \includegraphics[width=\columnwidth, trim={0.5cm 0 0.4cm 0},clip]{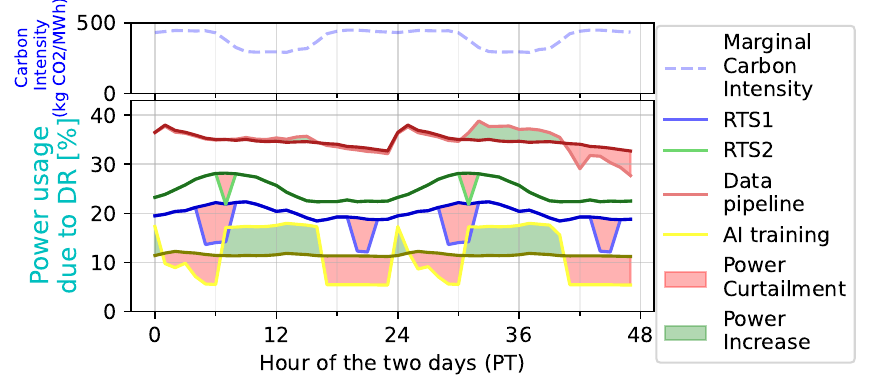} 
	\vspace{-.2in}
        \caption{Optimal power allocation of \ourmodel}
        \vspace{-.25in}
	\label{fig:ourmodel}
\end{figure}

Figure~\ref{fig:ourmodel} illustrates hourly marginal carbon intensity and power usage before and after \ourmodel. Lines present data for four workloads. Red and green areas between lines indicate negative and positive power adjustments, respectively. Adjustment $\mathbf{D}$ is calculated using CR1 and Optimization~\ref{eq:wo}. Collectively, the four workloads reduce operational carbon by 4.6\% and suffer a performance loss equivalent to a 4\% reduction in their power capacity.  

Real-time workloads reduce power usage (\textit{i.e.}, red adjustments) and degrade performance. \ig can tolerate reduced power and consistently curtails usage when carbon intensity is high, reducing carbon by 2.6\% and degrading performance by an equivalent of 2.9\% in power capacity. \web suffers higher performance losses from DR and curtails usage less often, reducing carbon by only 0.4\% and degrading performance by an equivalent of 0.7\% in power capacity.

Batch workloads defer power (\textit{i.e.}, red and green adjustments), shifting it to hours of low carbon intensity. AI training and data pipeline workloads defer jobs from the 6pm--8am window to the 8am--6pm window, reducing carbon by $1.2\%$ and $0.3\%$ while degrading performance by only 0.2\% and 0.3\%. 


Figure~\ref{fig:ourmodel} visualizes outcomes from policy CR1 with $\lambda=6.9$. The next section explores other policies and hyperparameters that could further reduce carbon or preserve performance. 



\subsection{Efficiency and Fairness}
\label{sec:efficiency}
\begin{figure}
	\centering
	\includegraphics[width=\columnwidth]{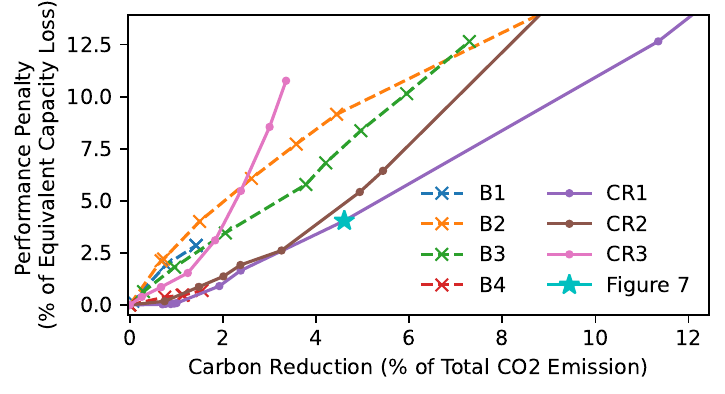}
   	\vspace*{-.25in}
\caption{{Performance and carbon trade-off of Carbon Responder (CR) and the baseline approaches (B\#).}}
	\label{fig:tradeoff}
        \vspace{-.2in}
\end{figure}

We evaluate the efficiency of carbon-informed DR based on its ability to reduce datacenter carbon while mitigating performance losses. By adjusting policies' hyperparameters (e.g., $\lambda$, cap\%, and $F$), we can obtain different DR outcomes and trade-offs between carbon and performance. 

Figure~\ref{fig:tradeoff} illustrates Pareto frontiers for baseline and \ourmodel policies. The x-axis represents carbon reductions, while the y-axis represents the total performance losses incurred, both as a percentage of numbers without DR. Upward-sloping curves indicate that as a policy reduces carbon more aggressively, performance losses increase. 

\textbf{Efficiency versus Fairness.} Policies with frontiers located on the lower right of the figure eliminate more carbon for the same level of performance loss, making them more efficient than those located on the upper left. For instance, when the performance loss ranges from 1\% to 5\%, CR1 eliminates twice as much carbon as baselines B1-B4.

CR1 is most efficient followed by CR2 and B4 (\textit{i.e.}, red line near origin). Optimization~\ref{eq:wo} and its weighted sum minimizes carbon under varied constraints on performance loss, thereby establishing an upper bound on efficiency. CR1 is globally optimal and efficient. Unfortunately, it is also unfair.

CR2 minimizes carbon under varied constraints on fairness, performing well initially but eventually suffering from an inevitable trade-off between efficiency and fairness.  CR3 suffers from a more severe version of this trade-off. Its Pareto frontier rises more quickly than CR2's because it uses decentralized incentives (\textit{i.e.}, tax and rebate), which allow workloads to make independent DR decisions but harm efficiency.



\textbf{Limits of Baseline Policies.}
When comparing the baseline policies to \ourmodel, we observe that B1, B2, and B3 are less efficient compared to CR1 and CR2. We analyze B1 without the batch preservation constraint. Otherwise, B1 would have terminated at the yellow start in Figure 8, indicating its inability to adjust power under the constraint. B2, despite having been designed to be performant, incurs a greater performance loss. This can be attributed to B2's additional objective of peak shaving and the constraint of batch preservation, which limits its efficiency. 
B4 is not at all effective at DR for carbon reduction. It only curtails batch workloads in accordance with their SLOs, resulting in negligible carbon reductions and performance losses because batch workloads without SLOs constitute a small share of our datacenter's total workload (Fig.~\ref{fig:dr}). 

\subsection{Service-Level Analysis}
\label{sec:perworkload}


\begin{figure}
	\centering
	\includegraphics[width=\columnwidth]{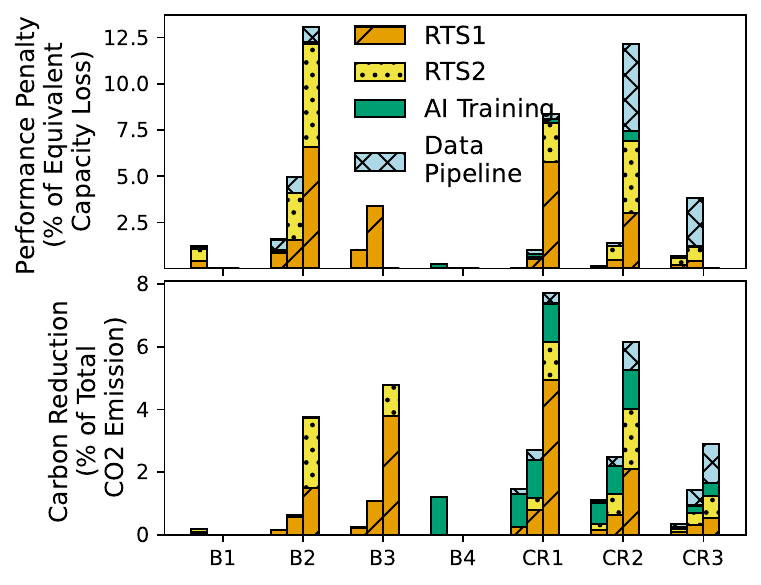}	
   	\vspace{-.25in}
\caption{{Breakdown of performance penalty and carbon reduction per workload for carbon reductions of 0.5, 2, and 8\%. A missing bar for B\# indicates it is incapable of achieving that level of carbon emission reduction.}}
        \vspace{-.2in}
\label{fig:bar}
\end{figure}

We assess the distribution of carbon reductions and performance losses across services and workloads. The upper subplot of Figure~\ref{fig:bar} details performance losses for each policy and service when the datacenter uses demand response to reduce carbon by 0.5\%, 2\%, and 8\%. Similarly, the lower subplot details carbon reductions. The absence of bars for B3, B4, and CR3 indicate their inability to achieve target carbon reductions or satisfy performance constraints. 



\textbf{Performance Loss.} 
In the upper subplot of Figure~\ref{fig:bar}, CR1 minimizes total performance penalty and its bars represent the most efficient distribution of those penalties. In contrast, CR2 emphasizes fairness and imposes performance penalties based on each workload's allocation of power capacity. Compared to CR1, CR2 allocates more penalties to \web than \ig because \web has a higher \pce. Note that AI training and data pipeline's penalties are not exactly proportional to their \pce due to constraints imposed by batch preservation. Finally, the upper subplot does not illustrate CR3's fairness because the policy defines fairness in terms of taxes and rebates rather than performance outcomes. 



\textbf{Carbon Reduction.}
The lower subplot of Figure~\ref{fig:bar} presents the breakdown of carbon reductions. Compared to CR1, CR2 distributes carbon reductions more evenly across workloads. Although CR3 reduces total carbon by less than the first two policies, it distributes those reductions in rough proportion to each workload's power capacity, thereby achieving the most equitable allocation of responsibilities across workloads. 

Baseline policies exhibit varying levels of efficiency and fairness. B1's proportional power capping is fair, in the distributions of both performance losses and carbon reductions, but also inefficient because power capping ignores time-varying carbon intensity. B2 is similarly inefficient, power capping only real-time workloads because their performance is relatively resilient to caps. On the other hand, B3 and B4 exclusively curtail either real-time or batch workloads, producing both unfair performance distributions and insufficient carbon reductions. 


\subsection{Fairness}
\label{sec:fairness}
\begin{figure}
	\centering
 \vspace{-.2in}
	\includegraphics[width=\columnwidth]{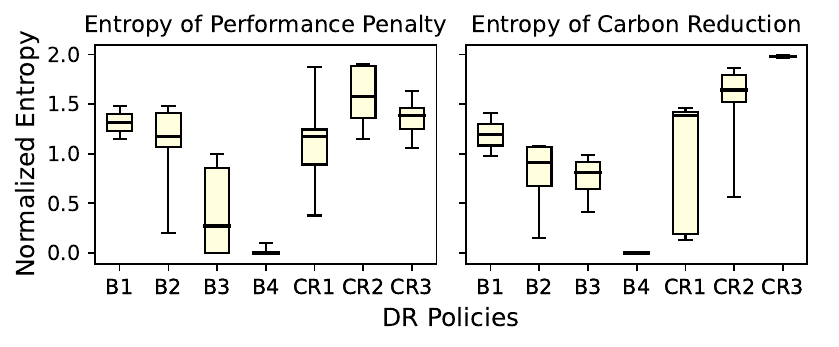}	
        
\caption{Normalized entropy of the performance penalties and carbon reduction across workloads of different policies.}
        \vspace{-.15in}
	\label{fig:entropy}
\end{figure}

We explicitly evaluate fairness by measuring Shannon entropy, which is $$-\sum_{i=1}^n p_i \log p_i$$ for a discrete distribution $p_i$. We consider four workloads ($n=4$) for which greater entropy indicates a more equitable distribution. Entropy has a maximum value of 2, which corresponds to a perfectly fair, uniform distribution.

We calculate entropy by scaling performance loss $C$ and carbon reduction $CF$ by each workload's \pce such that $p_i = {C_i}/{E_i} $ or $p_i = {CF_i}/{E_i}$. With this scaling, entropy aligns carbon breakdowns or performance losses and workloads' respective allocations of \pce. Specifically, when performance losses or carbon reductions are directly proportional to the \pce, entropy reaches the maximum value of 2.

Figure~\ref{fig:entropy} illustrates how closely performance losses and carbon reductions of workloads align with their \pce under each policy. The entropy distribution of each policy was calculated by varying their hyperparameters, following the same data points as shown in Figure~\ref{fig:tradeoff}. The Box-and-whisker plots present the 1st, 2nd, and 3rd quartiles for entropy, while the whiskers extend to the minimum and maximum values. A higher box height in the plot indicates a greater degree of fairness in the policy, while shorter boxes and whiskers indicate less variability and more consistent fairness results across different hyperparameters.

Policies B1, CR2, and CR3 emerge as the most fair policies due to their equal treatment of workloads. B1 applies proportional caps to all workloads, resulting in fair albeit inefficient DR. CR2 defines optimization constraints that achieves fair outcomes for performance losses. CR3 defines an incentive mechanism that rewards workloads for DR contributions, thereby achieving fair outcomes for carbon reductions. 

Policies that neglect fairness report low entropy. Policies B2 and CR1 optimize for performance and carbon without accounting for fairness. B3 and B4 discriminate between workloads and curtail power based on workload type.

\subsection{Increasing Potential in the Future}
\begin{figure}
 \vspace{-.1in}
		\centering\includegraphics[width=1\columnwidth]{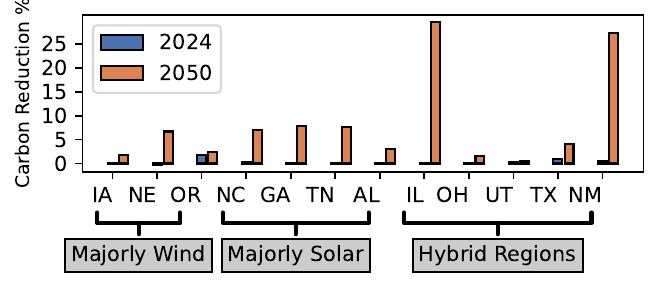}
    \vspace{-15pt}
	\caption{Projected gains from applying the load shifting from Figure 7 across all states in future scenarios~\cite{gagnon_cambium_2023}.}
 \vspace{-.1in}
	\label{fig:ca}
\end{figure}


\bhl{
Figure~{\ref{fig:ca}} demonstrates the significant potential of carbon-aware DR due to the growing variations of grid carbon intensities. Fixing the load adjustments as depicted in Figure~{\ref{fig:ourmodel}},} Figure~{\ref{fig:ca}} showcases the hypothetical carbon reduction of datacenters achieved thereby in 2024 and 2050. Each state exhibits different levels of potential benefits, influenced by their predicted carbon-free energy availability in 2024 and 2050\mbox{\cite{gagnon_cambium_2023}}. 
The bars in Figure~{\ref{fig:ca}} represent a lower bound on the carbon reduction for 2024 and 2050, \bhl{as it employs the load shift from today (Figure 7) rather than re-optimizing the load adjustments with CR.}
The substantial increase in carbon reduction from 2024 to 2050 can be attributed to the growing variations in carbon intensity across most states
As the deployment of solar energy continues to expand,
it leads to a remarkable surge in carbon reduction achieved through  \ourmodel.

%% file: 07_relatedwork.tex
\section{Discussion and Related Work}


We are motivated by prior work in supply function bidding \cite{johari_parameterized_2011, xu_demand_2016} as well as DR aggregators that both curtail and defer load \cite{talari_optimal_2019}. 

\textbf{Demand Response (DR).} Markets use monetary payments to align incentives between the grid and consumers \cite{wierman_opportunities_2014, liu_pricing_2014, zhou_when_2015}. Markets also incentivize co-located tenants to collaboratively reduce carbon \cite{islam_greencolo_2014, islam_paying_2015, islam_market_2016, ren_colocation_2014}. However, these works require well-defined markets with payment in dollars
at the grid level or within shared datacenters \cite{xu_two_2020, islam_greencolo_2014}. Within the datacenter fleet of a company, there is no prior work in modeling performance elasticity and defining DR payment. Datacenter carbon-aware DR adjusts hourly power capacity depending on the grid's carbon intensity \cite{katsak_greencassandra_2015, mahmud_online_2013, radovanovic_carbon-aware_2021}. Carbon-aware job schedulers can account for time-varying capacity constraints \cite{mencia_genetic_2017, gil-gala_genetic_2021, klusacek_scheduling_2021} or green energy availability \cite{goiri_greenhadoop_2012}. 
\bhl{Similarly, cloud applications can be provisioned with carbon-awareness by employing integer programming techniques that impose constraints on the usage of carbon-intensive grid energy} \cite{deng_policy_2012}.
However, these studies lack techniques for apportioning curtailments across heterogeneous batch and online workloads. 

\textbf{Performance/Power Management.} 
Prior studies model datacenter performance to calculate capacity and improve power efficiency \cite{chen_integrated_2010, gandhi_minimizing_2011, heo_optituner_2011, li_towards_2012, lin_dynamic_2011, meisner_power_2011, zhang_dynamic_2012}. 
Their models optimize processor and power utilization with performance targets, usually with SLAs based on queuing theory \cite{chen_integrated_2010, gandhi_minimizing_2011, heo_optituner_2011}. 
\bhl{However, power efficiency is not equivalent to carbon reduction because DR leverages the intermittency of renewable energy and shift power demand accordingly} {\cite{stewart_joules_2009}.}

\quad\textbf{Net Zero and Renewable Energy Credits.}
Today's datacenters often procure wind and solar projects, generating renewable energy credits (RECs) that allow them to offset their annual carbon footprints \cite{facebook21:tariffs, google13:ppa, microsoft21}. Such procurements facilitate claims of being \textit{100\% powered by renewable energy}. However, these claims, when examined on an hourly basis, reveal that a datacenter's energy consumption can frequently surpass the amount of procured renewables.
At these times, the datacenter's energy is only as green as the grid's broader carbon intensity. 

\textbf{Operational vs. Embodied Carbon Footprint.}
While embodied carbon dominates in battery-operated systems,
operational carbon remains significant in datacenters \cite{gupta_chasing_2021}.
According to the 2023 Meta and Google Sustainability Reports, without carbon offsetting (i.e., location-based approach), operational carbon comprises 41\% and 79\% of their total respective datacenter carbon footprints \cite{meta_2023_report, google_2023_report}. 
However, with annual renewable offsetting (i.e., market-based approach), operational carbon was greatly reduced to 1\% and 54\% respectively, making the carbon footprint predominantly embodied. With hourly renewable offsetting, we use the open-source Carbon Explorer framework and calculated a 40-75\% decrease (depending on the region characteristics) in operational carbon. Furthermore, since a portion of the embodied footprint is coming from electricity consumed during manufacturing, optimizing for both operational and embodied carbon is important.

\textbf{Embodied Carbon in Demand Response.}
Demand response, particularly the CR methodology, operates within current server capacities, preventing additional embodied carbon footprints. 
Even when demand response entails extra servers, the overall carbon footprint can still decrease. Carbon Explorer's recent findings affirm this, demonstrating that incorporating the embodied carbon of extra servers still results in a net reduction of carbon footprint in various regions \cite{acun23}.

%% file: 08_discussion.tex
\section{Conclusion}

We presented \ourmodel, a datacenter DR framework that addresses the challenges of carbon-informed DR in private datacenters supporting diverse workloads. We extend demand response to include both realtime and batch workloads with varying service level objectives, accounting for their sensitivity to power allocation. We introduce a family of performance-aware DR policies, exploring the trade-offs between efficiency, fairness, and carbon reduction. Experimental results demonstrate that the efficient CR policy achieves a remarkable carbon footprint reduction of 1.5x to 2x compared to baseline approaches, while the fair CR policies distribute responsibilities more equitably among workloads.

Beyond specific numbers for performance-carbon tradeoffs, CR's contribution is exploring the space of DR policies and highlighting inherent challenges overlooked in prior research. Without a thorough analysis of performance implications and a commitment to fairly distributing DR among diverse workloads, realizing DR in hyperscale datacenters remains a lofty ambition.
Though the carbon benefit from DR in a realistic datacenter setup may be modest today, it serves as a call to action. We need to design datacenters, hardware, and software with delay tolerance and energy proportionality in mind.




%% file: main.bbl
\begin{thebibliography}{10}
\providecommand{\url}[1]{#1}
\csname url@samestyle\endcsname
\providecommand{\newblock}{\relax}
\providecommand{\bibinfo}[2]{#2}
\providecommand{\BIBentrySTDinterwordspacing}{\spaceskip=0pt\relax}
\providecommand{\BIBentryALTinterwordstretchfactor}{4}
\providecommand{\BIBentryALTinterwordspacing}{\spaceskip=\fontdimen2\font plus
\BIBentryALTinterwordstretchfactor\fontdimen3\font minus
  \fontdimen4\font\relax}
\providecommand{\BIBforeignlanguage}[2]{{%
\expandafter\ifx\csname l@#1\endcsname\relax
\typeout{** WARNING: IEEEtranS.bst: No hyphenation pattern has been}%
\typeout{** loaded for the language `#1'. Using the pattern for}%
\typeout{** the default language instead.}%
\else
\language=\csname l@#1\endcsname
\fi
#2}}
\providecommand{\BIBdecl}{\relax}
\BIBdecl

\bibitem{california_iso}
\BIBentryALTinterwordspacing
``California {ISO} - {Managing} {Oversupply}.'' [Online]. Available:
  \url{https://www.caiso.com/informed/Pages/ManagingOversupply.aspx}
\BIBentrySTDinterwordspacing

\bibitem{acun23}
B.~Acun, B.~Lee, F.~Kazhamiaka, K.~Maeng, U.~Gupta, M.~Chakkaravarthy,
  D.~Brooks, and C.~Wu, ``Carbon explorer: A holistic framework for designing
  carbon aware datacenters,'' in \emph{Proc. Architectural Support for
  Programming Languages and Operating Systems (ASPLOS)}, 2023.

\bibitem{acun_carbon_2022}
\BIBentryALTinterwordspacing
B.~Acun, B.~Lee, F.~Kazhamiaka, A.~Sundarrajan, K.~Maeng, M.~Chakkaravarthy,
  D.~Brooks, and C.-J. Wu, ``Carbon {Dependencies} in {Datacenter} {Design} and
  {Management},'' in \emph{{HotCarbon}'22}, UC San Diego campus, La Jolla,
  California, 2022. [Online]. Available:
  \url{https://research.facebook.com/publications/carbon-dependencies-in-datacenter-design-and-management/}
\BIBentrySTDinterwordspacing

\bibitem{acun_holistic_2022}
\BIBentryALTinterwordspacing
B.~Acun, B.~Lee, K.~Maeng, M.~Chakkaravarthy, U.~Gupta, D.~Brooks, and C.-J.
  Wu, ``A {Holistic} {Approach} for {Designing} {Carbon} {Aware}
  {Datacenters},'' \emph{arXiv:2201.10036 [cs, eess]}, Jan. 2022, arXiv:
  2201.10036. [Online]. Available: \url{http://arxiv.org/abs/2201.10036}
\BIBentrySTDinterwordspacing

\bibitem{eia_outlook_23}
U.~E.~I. Administration", ``Annual energy outlook 2023,''
  \url{https://www.eia.gov/outlooks/aeo/narrative/index.php}, 2023.

\bibitem{amazon21}
{Amazon}, ``Delivering progress every day: {Amazon's} 2021 sustainability
  report,'' Tech. Rep., 2021.

\bibitem{buchanan_carbon-aware_2023}
\BIBentryALTinterwordspacing
W.~Buchanan, J.~Foxon, D.~Cooke, S.~Iyer, E.~Graham, B.~DeRusha, C.~Binder,
  K.~Chiu, H.~Richardson, V.~Knight, A.~Hussain, and N.~Mathews,
  ``\BIBforeignlanguage{en}{Carbon-aware computing: {Measuring} and reducing
  the carbon footprint associated with software in execution},''
  \emph{\BIBforeignlanguage{en}{Microsoft Switzerland News Center}}, Jan. 2023.
  [Online]. Available:
  \url{https://news.microsoft.com/de-ch/2023/01/10/carbon-aware-computing-whitepaper/}
\BIBentrySTDinterwordspacing

\bibitem{bunodiere_renewable_2020}
\BIBentryALTinterwordspacing
A.~Bunodiere and H.~S. Lee, ``\BIBforeignlanguage{en}{Renewable {Energy}
  {Curtailment}: {Prediction} {Using} a {Logic}-{Based} {Forecasting} {Method}
  and {Mitigation} {Measures} in {Kyushu}, {Japan}},''
  \emph{\BIBforeignlanguage{en}{Energies}}, vol.~13, no.~18, p. 4703, Jan.
  2020, number: 18 Publisher: Multidisciplinary Digital Publishing Institute.
  [Online]. Available: \url{https://www.mdpi.com/1996-1073/13/18/4703}
\BIBentrySTDinterwordspacing

\bibitem{chase01}
J.~Chase, D.~Anderson, P.~Thakar, A.~Vahdat, and R.~Doyle, ``Managing energy
  and server resources in hosting centers,'' in \emph{Proc Symposium on
  Operating System Principles (SOSP)}, 2001.

\bibitem{chen_quantifying_2006}
\BIBentryALTinterwordspacing
K.-T. Chen, C.-Y. Huang, P.~Huang, and C.-L. Lei, ``Quantifying {Skype} user
  satisfaction,'' \emph{ACM SIGCOMM Computer Communication Review}, vol.~36,
  no.~4, pp. 399--410, Aug. 2006. [Online]. Available:
  \url{https://dl.acm.org/doi/10.1145/1151659.1159959}
\BIBentrySTDinterwordspacing

\bibitem{chen_integrated_2010}
Y.~Chen, D.~Gmach, C.~Hyser, Z.~Wang, C.~Bash, C.~Hoover, and S.~Singhal,
  ``Integrated management of application performance, power and cooling in data
  centers,'' in \emph{2010 {IEEE} {Network} {Operations} and {Management}
  {Symposium} - {NOMS} 2010}, Apr. 2010, pp. 615--622, iSSN: 2374-9709.

\bibitem{cortez_resource_2017}
\BIBentryALTinterwordspacing
E.~Cortez, A.~Bonde, A.~Muzio, M.~Russinovich, M.~Fontoura, and R.~Bianchini,
  ``Resource {Central}: {Understanding} and {Predicting} {Workloads} for
  {Improved} {Resource} {Management} in {Large} {Cloud} {Platforms},'' in
  \emph{Proceedings of the 26th {Symposium} on {Operating} {Systems}
  {Principles}}, ser. {SOSP} '17.\hskip 1em plus 0.5em minus 0.4em\relax New
  York, NY, USA: Association for Computing Machinery, 2017, pp. 153--167.
  [Online]. Available: \url{http://doi.org/10.1145/3132747.3132772}
\BIBentrySTDinterwordspacing

\bibitem{deng_policy_2012}
\BIBentryALTinterwordspacing
N.~Deng, C.~Stewart, D.~Gmach, and M.~Arlitt, ``Policy and mechanism for
  carbon-aware cloud applications,'' in \emph{2012 {IEEE} {Network}
  {Operations} and {Management} {Symposium}}, Apr. 2012, pp. 590--594, iSSN:
  2374-9709. [Online]. Available:
  \url{https://ieeexplore.ieee.org/document/6211963}
\BIBentrySTDinterwordspacing

\bibitem{deng_concentrating_2011}
\BIBentryALTinterwordspacing
N.~Deng, C.~Stewart, and J.~Li, ``Concentrating renewable energy in grid-tied
  datacenters,'' in \emph{Proceedings of the 2011 {IEEE} {International}
  {Symposium} on {Sustainable} {Systems} and {Technology}}, May 2011, pp. 1--6,
  iSSN: 2378-7260. [Online]. Available:
  \url{https://ieeexplore.ieee.org/abstract/document/5936855}
\BIBentrySTDinterwordspacing

\bibitem{facebook21:tariffs}
{Facebook}, ``Advancing renewable energy through green tariffs,'' Tech. Rep.,
  2021.

\bibitem{fan_power_2007}
\BIBentryALTinterwordspacing
X.~Fan, W.-D. Weber, and L.~A. Barroso, ``Power provisioning for a
  warehouse-sized computer,'' \emph{ACM SIGARCH Computer Architecture News},
  vol.~35, no.~2, pp. 13--23, Jun. 2007. [Online]. Available:
  \url{https://doi.org/10.1145/1273440.1250665}
\BIBentrySTDinterwordspacing

\bibitem{gagnon_cambium_2023}
\BIBentryALTinterwordspacing
P.~Gagnon, B.~Cowiestoll, and M.~Schwarz, ``Cambium 2022 {Scenario}
  {Descriptions} and {Documentation},'' National Renewable Energy Lab.(NREL),
  Golden, CO (United States), Tech. Rep., 2023. [Online]. Available:
  \url{https://www.nrel.gov/analysis/cambium.html}
\BIBentrySTDinterwordspacing

\bibitem{gagnon2022long}
P.~Gagnon, E.~Hale, and W.~Cole, ``Long-run marginal emission rates for
  electricity-workbooks for 2021 cambium data,'' National Renewable Energy
  Laboratory-Data (NREL-DATA), Golden, CO: National Renewable Energy
  Laboratory, Tech. Rep., 2022.

\bibitem{gandhi_minimizing_2011}
A.~Gandhi, Y.~Chen, D.~Gmach, M.~Arlitt, and M.~Marwah, ``Minimizing data
  center {SLA} violations and power consumption via hybrid resource
  provisioning,'' in \emph{2011 {International} {Green} {Computing}
  {Conference} and {Workshops}}, Jul. 2011, pp. 1--8.

\bibitem{gil-gala_genetic_2021}
\BIBentryALTinterwordspacing
F.~J. Gil-Gala, M.~R. Sierra, C.~Mencía, and R.~Varela,
  ``\BIBforeignlanguage{en}{Genetic programming with local search to evolve
  priority rules for scheduling jobs on a machine with time-varying
  capacity},'' \emph{\BIBforeignlanguage{en}{Swarm and Evolutionary
  Computation}}, vol.~66, p. 100944, Oct. 2021. [Online]. Available:
  \url{https://www.sciencedirect.com/science/article/pii/S2210650221001061}
\BIBentrySTDinterwordspacing

\bibitem{goiri_greenhadoop_2012}
\BIBentryALTinterwordspacing
I.~n. Goiri, K.~Le, T.~D. Nguyen, J.~Guitart, J.~Torres, and R.~Bianchini,
  ``{GreenHadoop}: leveraging green energy in data-processing frameworks,'' in
  \emph{Proceedings of the 7th {ACM} european conference on {Computer}
  {Systems}}, ser. {EuroSys} '12.\hskip 1em plus 0.5em minus 0.4em\relax New
  York, NY, USA: Association for Computing Machinery, 2012, pp. 57--70.
  [Online]. Available: \url{http://doi.org/10.1145/2168836.2168843}
\BIBentrySTDinterwordspacing

\bibitem{goiri_greenslot_2011}
{\'I}.~Goiri, K.~Le, M.~E. Haque, R.~Beauchea, T.~D. Nguyen, J.~Guitart,
  J.~Torres, and R.~Bianchini, ``Greenslot: scheduling energy consumption in
  green datacenters,'' in \emph{Proceedings of 2011 International Conference
  for High Performance Computing, Networking, Storage and Analysis}, 2011, pp.
  1--11.

\bibitem{google_2023_report}
\BIBentryALTinterwordspacing
Google, ``\BIBforeignlanguage{en-US}{2023 {Environmental} {Report}}.''
  [Online]. Available:
  \url{https://sustainability.google/reports/google-2023-environmental-report/}
\BIBentrySTDinterwordspacing

\bibitem{google13:ppa}
{Google}, ``{Google's} green {PPAs}: {What}, how and why,'' Tech. Rep., 2013.

\bibitem{google21}
{Google}, ``Environmental report,'' Tech. Rep., 2021.

\bibitem{govindan_benefits_2011}
\BIBentryALTinterwordspacing
S.~Govindan, A.~Sivasubramaniam, and B.~Urgaonkar, ``Benefits and limitations
  of tapping into stored energy for datacenters,'' in \emph{Proceedings of the
  38th annual international symposium on {Computer} architecture}, ser. {ISCA}
  '11.\hskip 1em plus 0.5em minus 0.4em\relax New York, NY, USA: Association
  for Computing Machinery, Jun. 2011, pp. 341--352. [Online]. Available:
  \url{https://doi.org/10.1145/2000064.2000105}
\BIBentrySTDinterwordspacing

\bibitem{graham_optimization_1979}
\BIBentryALTinterwordspacing
R.~L. Graham, E.~L. Lawler, J.~K. Lenstra, and A.~H. G.~R. Kan,
  ``\BIBforeignlanguage{en}{Optimization and {Approximation} in {Deterministic}
  {Sequencing} and {Scheduling}: a {Survey}},'' in
  \emph{\BIBforeignlanguage{en}{Annals of {Discrete} {Mathematics}}}, ser.
  Discrete {Optimization} {II}, P.~L. Hammer, E.~L. Johnson, and B.~H. Korte,
  Eds.\hskip 1em plus 0.5em minus 0.4em\relax Elsevier, Jan. 1979, vol.~5, pp.
  287--326. [Online]. Available:
  \url{https://www.sciencedirect.com/science/article/pii/S016750600870356X}
\BIBentrySTDinterwordspacing

\bibitem{guevara13}
M.~Guevara, B.~Lubin, and B.~Lee, ``Navigating heterogeneous processors with
  market mechanisms,'' in \emph{Proc. Symposium on High-Performance Computer
  Architecture (HPCA)}, 2013.

\bibitem{gupta_chasing_2021}
U.~Gupta, Y.~G. Kim, S.~Lee, J.~Tse, H.-H.~S. Lee, G.-Y. Wei, D.~Brooks, and
  C.-J. Wu, ``Chasing {Carbon}: {The} {Elusive} {Environmental} {Footprint} of
  {Computing},'' in \emph{2021 {IEEE} {International} {Symposium} on
  {High}-{Performance} {Computer} {Architecture} ({HPCA})}, Feb. 2021, pp.
  854--867, iSSN: 2378-203X.

\bibitem{he_zeta_2012}
\BIBentryALTinterwordspacing
Y.~He, S.~Elnikety, J.~Larus, and C.~Yan, ``Zeta: scheduling interactive
  services with partial execution,'' in \emph{Proceedings of the {Third} {ACM}
  {Symposium} on {Cloud} {Computing}}, ser. {SoCC} '12.\hskip 1em plus 0.5em
  minus 0.4em\relax New York, NY, USA: Association for Computing Machinery,
  Oct. 2012, pp. 1--14. [Online]. Available:
  \url{https://dl.acm.org/doi/10.1145/2391229.2391241}
\BIBentrySTDinterwordspacing

\bibitem{heo_optituner_2011}
J.~Heo, P.~Jayachandran, I.~Shin, D.~Wang, T.~Abdelzaher, and X.~Liu,
  ``{OptiTuner}: {On} {Performance} {Composition} and {Server} {Farm} {Energy}
  {Minimization} {Application},'' \emph{IEEE Transactions on Parallel and
  Distributed Systems}, vol.~22, no.~11, pp. 1871--1878, Nov. 2011, conference
  Name: IEEE Transactions on Parallel and Distributed Systems.

\bibitem{islam_paying_2015}
M.~A. Islam, H.~Mahmud, S.~Ren, and X.~Wang, ``Paying to save: {Reducing} cost
  of colocation data center via rewards,'' in \emph{2015 {IEEE} 21st
  {International} {Symposium} on {High} {Performance} {Computer} {Architecture}
  ({HPCA})}, Feb. 2015, pp. 235--245, iSSN: 2378-203X.

\bibitem{islam_greencolo_2014}
M.~A. Islam, S.~Ren, and X.~Wang, ``{GreenColo}: {A} novel incentive mechanism
  for minimizing carbon footprint in colocation data center,'' in
  \emph{International {Green} {Computing} {Conference}}, Nov. 2014, pp. 1--8.

\bibitem{islam_market_2016}
M.~A. Islam, X.~Ren, S.~Ren, A.~Wierman, and X.~Wang, ``A market approach for
  handling power emergencies in multi-tenant data center,'' in \emph{2016
  {IEEE} {International} {Symposium} on {High} {Performance} {Computer}
  {Architecture} ({HPCA})}, Mar. 2016, pp. 432--443, iSSN: 2378-203X.

\bibitem{johari_parameterized_2011}
\BIBentryALTinterwordspacing
R.~Johari and J.~N. Tsitsiklis, ``Parameterized {Supply} {Function} {Bidding}:
  {Equilibrium} and {Efficiency},'' \emph{Operations Research}, vol.~59, no.~5,
  pp. 1079--1089, Oct. 2011, publisher: INFORMS. [Online]. Available:
  \url{https://pubsonline.informs.org/doi/abs/10.1287/opre.1110.0980}
\BIBentrySTDinterwordspacing

\bibitem{katsak_greencassandra_2015}
W.~Katsak, I.~n. Goiri, R.~Bianchini, and T.~D. Nguyen, ``{GreenCassandra}:
  {Using} renewable energy in distributed structured storage systems,'' in
  \emph{2015 {Sixth} {International} {Green} and {Sustainable} {Computing}
  {Conference} ({IGSC})}, Dec. 2015, pp. 1--8.

\bibitem{koulamas_total_1994}
\BIBentryALTinterwordspacing
C.~Koulamas, ``The {Total} {Tardiness} {Problem}: {Review} and {Extensions},''
  \emph{Operations Research}, vol.~42, no.~6, pp. 1025--1041, Dec. 1994,
  publisher: INFORMS. [Online]. Available:
  \url{https://pubsonline.informs.org/doi/abs/10.1287/opre.42.6.1025}
\BIBentrySTDinterwordspacing

\bibitem{krioukov_design_2012}
\BIBentryALTinterwordspacing
A.~Krioukov, S.~Alspaugh, P.~Mohan, S.~Dawson-Haggerty, D.~E. Culler, and R.~H.
  Katz, ``Design and {Evaluation} of an {Energy} {Agile} {Computing}
  {Cluster},'' EECS Department, University of California, Berkeley, Tech. Rep.
  UCB/EECS-2012-13, Jan. 2012. [Online]. Available:
  \url{http://www2.eecs.berkeley.edu/Pubs/TechRpts/2012/EECS-2012-13.html}
\BIBentrySTDinterwordspacing

\bibitem{kumbhare_prediction-based_2021}
\BIBentryALTinterwordspacing
A.~G. Kumbhare, R.~Azimi, I.~Manousakis, A.~Bonde, F.~Frujeri, N.~Mahalingam,
  P.~A. Misra, S.~A. Javadi, B.~Schroeder, M.~Fontoura, and R.~Bianchini,
  ``\BIBforeignlanguage{en}{\{{Prediction}-{Based}\} {Power} {Oversubscription}
  in {Cloud} {Platforms}},'' 2021, pp. 473--487. [Online]. Available:
  \url{https://www.usenix.org/conference/atc21/presentation/kumbhare}
\BIBentrySTDinterwordspacing

\bibitem{li_towards_2012}
J.~Li, Z.~Li, K.~Ren, and X.~Liu, ``Towards {Optimal} {Electric} {Demand}
  {Management} for {Internet} {Data} {Centers},'' \emph{IEEE Transactions on
  Smart Grid}, vol.~3, no.~1, pp. 183--192, Mar. 2012, conference Name: IEEE
  Transactions on Smart Grid.

\bibitem{lin_dynamic_2011}
M.~Lin, A.~Wierman, L.~L.~H. Andrew, and E.~Thereska, ``Dynamic right-sizing
  for power-proportional data centers,'' in \emph{2011 {Proceedings} {IEEE}
  {INFOCOM}}, Apr. 2011, pp. 1098--1106, iSSN: 0743-166X.

\bibitem{liu_renewable_2012}
\BIBentryALTinterwordspacing
Z.~Liu, Y.~Chen, C.~Bash, A.~Wierman, D.~Gmach, Z.~Wang, M.~Marwah, and
  C.~Hyser, ``Renewable and cooling aware workload management for sustainable
  data centers,'' \emph{ACM SIGMETRICS Performance Evaluation Review}, vol.~40,
  no.~1, pp. 175--186, 2012. [Online]. Available:
  \url{http://doi.org/10.1145/2318857.2254779}
\BIBentrySTDinterwordspacing

\bibitem{liu_pricing_2014}
\BIBentryALTinterwordspacing
Z.~Liu, I.~Liu, S.~Low, and A.~Wierman, ``Pricing data center demand
  response,'' in \emph{The 2014 {ACM} international conference on {Measurement}
  and modeling of computer systems}, ser. {SIGMETRICS} '14.\hskip 1em plus
  0.5em minus 0.4em\relax New York, NY, USA: Association for Computing
  Machinery, Jun. 2014, pp. 111--123. [Online]. Available:
  \url{https://doi.org/10.1145/2591971.2592004}
\BIBentrySTDinterwordspacing

\bibitem{mahmud_online_2013}
\BIBentryALTinterwordspacing
A.~H. Mahmud and S.~Ren, ``Online capacity provisioning for carbon-neutral data
  center with demand-responsive electricity prices,'' \emph{ACM SIGMETRICS
  Performance Evaluation Review}, vol.~41, no.~2, pp. 26--37, 2013. [Online].
  Available: \url{http://doi.org/10.1145/2518025.2518029}
\BIBentrySTDinterwordspacing

\bibitem{meisner_power_2011}
\BIBentryALTinterwordspacing
D.~Meisner, C.~M. Sadler, L.~A. Barroso, W.-D. Weber, and T.~F. Wenisch,
  ``Power management of online data-intensive services,'' in \emph{Proceedings
  of the 38th annual international symposium on {Computer} architecture}, ser.
  {ISCA} '11.\hskip 1em plus 0.5em minus 0.4em\relax New York, NY, USA:
  Association for Computing Machinery, 2011, pp. 319--330. [Online]. Available:
  \url{http://doi.org/10.1145/2000064.2000103}
\BIBentrySTDinterwordspacing

\bibitem{mencia_genetic_2017}
C.~Mencía, M.~R. Sierra, R.~Mencía, and R.~Varela,
  ``\BIBforeignlanguage{en}{Genetic {Algorithm} for {Scheduling} {Charging}
  {Times} of {Electric} {Vehicles} {Subject} to {Time} {Dependent} {Power}
  {Availability}},'' in \emph{\BIBforeignlanguage{en}{Natural and {Artificial}
  {Computation} for {Biomedicine} and {Neuroscience}}}, ser. Lecture {Notes} in
  {Computer} {Science}, J.~M. Ferrández~Vicente, J.~R. Álvarez Sánchez,
  F.~de~la Paz~López, J.~Toledo~Moreo, and H.~Adeli, Eds.\hskip 1em plus 0.5em
  minus 0.4em\relax Cham: Springer International Publishing, 2017, pp.
  160--169.

\bibitem{meta_2023_report}
\BIBentryALTinterwordspacing
Meta, ``\BIBforeignlanguage{en-US}{2023 {Sustainability} {Report}}.'' [Online].
  Available: \url{https://sustainability.fb.com/2023-sustainability-report/}
\BIBentrySTDinterwordspacing

\bibitem{meta21}
{Meta}, ``Sustainability report,'' Tech. Rep., 2021.

\bibitem{microsoft21}
{Microsoft}, ``Environmental sustainability report,'' Tech. Rep., 2021.

\bibitem{radovanovic_carbon-aware_2021}
\BIBentryALTinterwordspacing
A.~Radovanovic, R.~Koningstein, I.~Schneider, B.~Chen, A.~Duarte, B.~Roy,
  D.~Xiao, M.~Haridasan, P.~Hung, N.~Care, S.~Talukdar, E.~Mullen, K.~Smith,
  M.~Cottman, and W.~Cirne, ``Carbon-{Aware} {Computing} for {Datacenters},''
  \emph{arXiv:2106.11750 [cs, eess]}, Jun. 2021, arXiv: 2106.11750. [Online].
  Available: \url{http://arxiv.org/abs/2106.11750}
\BIBentrySTDinterwordspacing

\bibitem{reddy_optimizing_2017}
\BIBentryALTinterwordspacing
S.~S. Reddy, ``\BIBforeignlanguage{en}{Optimizing energy and demand response
  programs using multi-objective optimization},''
  \emph{\BIBforeignlanguage{en}{Electrical Engineering}}, vol.~99, no.~1, pp.
  397--406, Mar. 2017. [Online]. Available:
  \url{https://doi.org/10.1007/s00202-016-0438-6}
\BIBentrySTDinterwordspacing

\bibitem{ren_colocation_2014}
\BIBentryALTinterwordspacing
S.~Ren and M.~A. Islam, ``\BIBforeignlanguage{en}{Colocation {Demand}
  {Response}: {Why} {Do} {I} {Turn} {Off} {My} {Servers}?}'' 2014, pp.
  201--208. [Online]. Available:
  \url{https://www.usenix.org/conference/icac14/technical-sessions/presentation/ren}
\BIBentrySTDinterwordspacing

\bibitem{sakalkar_data_2020}
\BIBentryALTinterwordspacing
V.~Sakalkar, V.~Kontorinis, D.~Landhuis, S.~Li, D.~De~Ronde, T.~Blooming,
  A.~Ramesh, J.~Kennedy, C.~Malone, J.~Clidaras, and P.~Ranganathan, ``Data
  {Center} {Power} {Oversubscription} with a {Medium} {Voltage} {Power} {Plane}
  and {Priority}-{Aware} {Capping},'' in \emph{Proceedings of the
  {Twenty}-{Fifth} {International} {Conference} on {Architectural} {Support}
  for {Programming} {Languages} and {Operating} {Systems}}, ser. {ASPLOS}
  '20.\hskip 1em plus 0.5em minus 0.4em\relax New York, NY, USA: Association
  for Computing Machinery, Mar. 2020, pp. 497--511. [Online]. Available:
  \url{https://dl.acm.org/doi/10.1145/3373376.3378533}
\BIBentrySTDinterwordspacing

\bibitem{stewart_joules_2009}
\BIBentryALTinterwordspacing
C.~Stewart and K.~Shen, ``Some joules are more precious than others: {Managing}
  renewable energy in the datacenter,'' in \emph{Proceedings of the workshop on
  power aware computing and systems}, 2009, pp. 15--19. [Online]. Available:
  \url{https://www.cs.rochester.edu/~kshen/papers/hotpower2009.pdf}
\BIBentrySTDinterwordspacing

\bibitem{talari_optimal_2019}
S.~Talari, M.~Shafie-khah, F.~Wang, J.~Aghaei, and J.~P.~S. Catalão, ``Optimal
  {Scheduling} of {Demand} {Response} in {Pre}-{Emptive} {Markets} {Based} on
  {Stochastic} {Bilevel} {Programming} {Method},'' \emph{IEEE Transactions on
  Industrial Electronics}, vol.~66, no.~2, pp. 1453--1464, 2019, conference
  Name: IEEE Transactions on Industrial Electronics.

\bibitem{tirmazi_borg_2020}
\BIBentryALTinterwordspacing
M.~Tirmazi, A.~Barker, N.~Deng, M.~E. Haque, Z.~G. Qin, S.~Hand,
  M.~Harchol-Balter, and J.~Wilkes, ``Borg: the next generation,'' in
  \emph{Proceedings of the {Fifteenth} {European} {Conference} on {Computer}
  {Systems}}, ser. {EuroSys} '20.\hskip 1em plus 0.5em minus 0.4em\relax New
  York, NY, USA: Association for Computing Machinery, 2020, pp. 1--14.
  [Online]. Available: \url{http://doi.org/10.1145/3342195.3387517}
\BIBentrySTDinterwordspacing

\bibitem{watttime_mer}
WattTime, ``Marginal emissions methodology,''
  \url{https://www.watttime.org/marginal-emissions-methodology/}, 2023.

\bibitem{wierman_opportunities_2014}
\BIBentryALTinterwordspacing
A.~Wierman, Z.~Liu, I.~Liu, and H.~Mohsenian-Rad,
  ``\BIBforeignlanguage{en}{Opportunities and challenges for data center demand
  response},'' in \emph{\BIBforeignlanguage{en}{International {Green}
  {Computing} {Conference}}}.\hskip 1em plus 0.5em minus 0.4em\relax DALLAS,
  TX, USA: IEEE, Nov. 2014, pp. 1--10. [Online]. Available:
  \url{http://ieeexplore.ieee.org/document/7039172/}
\BIBentrySTDinterwordspacing

\bibitem{wu_dynamo_2016}
\BIBentryALTinterwordspacing
Q.~Wu, Q.~Deng, L.~Ganesh, C.-H. Hsu, Y.~Jin, S.~Kumar, B.~Li, J.~Meza, and
  Y.~J. Song, ``Dynamo: facebook's data center-wide power management system,''
  \emph{ACM SIGARCH Computer Architecture News}, vol.~44, no.~3, pp. 469--480,
  2016. [Online]. Available: \url{http://doi.org/10.1145/3007787.3001187}
\BIBentrySTDinterwordspacing

\bibitem{xu_two_2020}
H.~Xu, X.~Jin, F.~Kong, and Q.~Deng, ``Two {Level} {Colocation} {Demand}
  {Response} with {Renewable} {Energy},'' \emph{IEEE Transactions on
  Sustainable Computing}, vol.~5, no.~1, pp. 147--159, 2020, conference Name:
  IEEE Transactions on Sustainable Computing.

\bibitem{xu_demand_2016}
Y.~Xu, N.~Li, and S.~H. Low, ``Demand {Response} {With} {Capacity}
  {Constrained} {Supply} {Function} {Bidding},'' \emph{IEEE Transactions on
  Power Systems}, vol.~31, no.~2, pp. 1377--1394, Mar. 2016, conference Name:
  IEEE Transactions on Power Systems.

\bibitem{zahedi18}
S.~M. Zahedi, Q.~Llull, and B.~Lee, ``Amdahl’s law in the datacenter era: {A}
  market for fair processor allocation,'' in \emph{HPCA}, 2018.

\bibitem{klusacek_scheduling_2021}
\BIBentryALTinterwordspacing
C.~Zhang and A.~A. Chien, ``\BIBforeignlanguage{en}{Scheduling {Challenges} for
  {Variable} {Capacity} {Resources}},'' in \emph{\BIBforeignlanguage{en}{Job
  {Scheduling} {Strategies} for {Parallel} {Processing}}}, D.~Klusáček,
  W.~Cirne, and G.~P. Rodrigo, Eds.\hskip 1em plus 0.5em minus 0.4em\relax
  Cham: Springer International Publishing, 2021, vol. 12985, pp. 190--209,
  series Title: Lecture Notes in Computer Science. [Online]. Available:
  \url{https://link.springer.com/10.1007/978-3-030-88224-2_10}
\BIBentrySTDinterwordspacing

\bibitem{zhang_dynamic_2012}
\BIBentryALTinterwordspacing
Q.~Zhang, M.~F. Zhani, S.~Zhang, Q.~Zhu, R.~Boutaba, and J.~L. Hellerstein,
  ``Dynamic energy-aware capacity provisioning for cloud computing
  environments,'' in \emph{Proceedings of the 9th international conference on
  {Autonomic} computing}, ser. {ICAC} '12.\hskip 1em plus 0.5em minus
  0.4em\relax New York, NY, USA: Association for Computing Machinery, 2012, pp.
  145--154. [Online]. Available: \url{http://doi.org/10.1145/2371536.2371562}
\BIBentrySTDinterwordspacing

\bibitem{zhang_hpc_2022}
Y.~Zhang, D.~C. Wilson, I.~C. Paschalidis, and A.~K. Coskun, ``{HPC} {Data}
  {Center} {Participation} in {Demand} {Response}: {An} {Adaptive} {Policy}
  {With} {QoS} {Assurance},'' \emph{IEEE Transactions on Sustainable
  Computing}, vol.~7, no.~1, pp. 157--171, Jan. 2022, conference Name: IEEE
  Transactions on Sustainable Computing.

\bibitem{zhou_when_2015}
Z.~Zhou, F.~Liu, Z.~Li, and H.~Jin, ``When smart grid meets geo-distributed
  cloud: {An} auction approach to datacenter demand response,'' in \emph{2015
  {IEEE} {Conference} on {Computer} {Communications} ({INFOCOM})}, Apr. 2015,
  pp. 2650--2658, iSSN: 0743-166X.

\end{thebibliography}
